\documentclass[lettersize,journal]{IEEEtran}
\usepackage{amsmath,amsfonts}
\usepackage{algorithmic}
\usepackage{algorithm}
\usepackage{array}
\usepackage{textcomp}
\usepackage{amssymb}
\usepackage{stfloats}
\usepackage{url}
\usepackage{bm}

\usepackage{verbatim}

\usepackage{graphicx}
\usepackage{subcaption}
\usepackage{tabularx}

\usepackage{cite}
\usepackage{color}
\usepackage{hyperref}
\hypersetup{
 colorlinks=true, 
 linkcolor=green,
 citecolor=blue,
 filecolor=red,
 urlcolor=blue,
 }
\usepackage{booktabs}

\usepackage{amsthm}
\newtheorem{remark}{Remark}

\newtheorem{lemma}{Lemma}

\usepackage{gensymb}
\allowdisplaybreaks[1] 

\usepackage[dvipsnames]{xcolor}

\graphicspath{{Images/}}

\begin{document}

\title{Uninformed-to-Informed Estimation: A Ping-Pong Positioning Method for Multi-user Wideband MmWave Systems}

\author{Lin~Guo,~Tiejun~Lv,~Yashuai~Cao,~and~Mugen~Peng
\thanks{Manuscript received 14 January 2025; revised 26 March 2025, and 24 July 2025; accepted 23 August 2025. This paper was supported in part by the National Natural Science Foundation of China under No. 62271068, and the Beijing Natural Science Foundation under Grant No. L222046. (\emph{corresponding author: Tiejun Lv}.)}

\thanks{L. Guo, T. Lv and M. Peng are with the School of Information and Communication Engineering, Beijing University of Posts and Telecommunications (BUPT), Beijing 100876, China (e-mail: \{gabrielle, lvtiejun, pmg\}@bupt.edu.cn).}
\thanks{Y. Cao is with the Department of Electronic Engineering, Tsinghua University, Beijing 100084, China, also with the State Key Laboratory of Space Network Communications, Beijing 100084, China, and also with the Beijing National Research Center for Information Science and Technology (BNRist), Beijing 100084, China (e-mail: caoys@tsinghua.edu.cn).}
}


\maketitle

\begin{abstract}
To enhance the positioning and tracking performance of dynamic user equipment (UE) in wideband millimeter-wave (mmWave) systems, we propose a novel positioning error lower bound (PELB)-driven ping-pong positioning framework, where the base station (BS) and UE alternately transmit and receive adaptive beamforming signals for positioning. All beamformers are scheduled based on the locally evaluated PELB. In this framework, we exploit multi-dimensional information fusion to assist in positioning. Firstly, a multi-subcarrier collaborative positioning error lower bound (MSCPEB) is proposed to evaluate the positioning error limits of wideband mmWave systems, which quantifies the contribution of all subcarriers to positioning accuracy. Moreover, we prove that the MSCPEB does not exceed the arithmetic mean of the PELBs of the individual subcarriers.
Subsequently, we develop an alternating optimization (AO) algorithm to optimize the hybrid beamformers targeted for MSCPEB minimization. By convexifying this problem, closed-form solutions of beamformers are derived. 
Finally, we develop a multipath collaborative positioning method that quantifies the impact of path reliability on positioning accuracy, with a closed-form solution for user position derived. The proposed method does not rely on path resolution and traditional triangular relationships.
Numerical results validate that the proposed method improves estimation accuracy by at least 16\% compared to potential schemes without optimized beam configurations, while requiring only approximately one-quarter of the slot resources.
\end{abstract}

\begin{IEEEkeywords}
MmWave positioning, Cram\'{e}r-Rao lower bound, hybrid beamforming, multi-dimensional information fusion.
\end{IEEEkeywords}

\section{Introduction}
\IEEEPARstart{H}{igh}-precision positioning is crucial for vertical applications, such as personal navigation, autonomous driving~\cite{8766143}, smart factories~\cite{9599656}, and virtual reality~\cite{9112752}. Beyond these applications, it holds significant promise for enhancing network capabilities in location-aware communications~\cite{7984759}.
As a prevailing positioning technology, the global positioning system (GPS) is limited by meter-level accuracy and high latency~\cite{8226757}.
In the context of fifth-generation (5G) wireless networks, millimeter-wave (mmWave) communications have garnered substantial attention due to their high transmission rates, spatial and temporal resolution within the 30--300 GHz frequency spectrum~\cite{10193776, 10054381}.
Combined with massive multiple-input multiple-output (MIMO), mmWave systems offer remarkable channel capacity, high beam directionality, and precise channel parameter estimation~\cite{8207426, 6834753, 10422712}.
Moreover, mmWave MIMO not only enables high-delay resolution with large available bandwidths, but also offers superior angular resolution. Consequently, mmWave MIMO has emerged as a promising positioning solution in 5G and beyond~\cite{3gpp2019nr, 9665433}.
Current studies on mmWave positioning primarily rely on spatial spectrum estimation (or channel parameter estimation)~\cite{9815098}. However, these methods do not fully take advantage of the benefits of large bandwidth and massive antenna arrays. In practice, the positioning performance limit is constrained by the signal processing operations of mmWave MIMO systems. Existing parameter estimation-based positioning techniques directly utilize signal operations targeted for communication-only performance criteria and do not explore optimal resource configurations that could enhance positioning performance.
As a result, these algorithms encounter a performance bottleneck in mmWave positioning. 
To conquer this issue, optimizing signal operations to enhance positioning performance limits is essential for developing fundamentally high-precision positioning solutions.

\subsection{Related Work}
Several studies have identified that exploring the positioning performance limit over transceiving resources can accommodate the wireless positioning demands~\cite{6799306,10323191}. Here, the positioning error lower bound (PELB) is typically derived to enhance positioning accuracy from the perspective of radio resources instead of the algorithm.
For instance, the authors of~\cite{6799306} examined the impact of beamforming strategies on mmWave position performance limits in distributed antenna systems. Similarly, the authors of~\cite{10323191} optimized the satellite beamforming to minimize the Cram\'{e}r-Rao lower bound (CRLB). Then, the CRLB minimization problem was equivalently cast as a signal-to-interference-plus-noise ratio (SINR) maximization problem. However, these beam scheduling methods for positioning require perfect channel state information (CSI), which is often not available in wireless positioning applications.

Some existing work has studied the lower bound of position estimation error for various scenarios. To optimize the PELB, a sensing-assisted beamforming strategy was proposed in~\cite{10547696,10579914, 9578931}.
The CRLB was utilized to evaluate the PELB of the proposed mmWave positioning methods in~\cite{9566601, 8709732}. However, these methods are only applicable to narrowband or single-user scenarios, without considering the rich additional information available in wide bandwidths. In~\cite{hua}, PELB was further optimized through hybrid analog and digital precoding, specifically tailored for single-antenna users. Additionally, a wideband localization scenario was studied in~\cite{10050406}, where two maximum likelihood (ML)-based multi-target localization schemes were designed. While this study addressed a wideband scenario, it only accounts for a single subcarrier.


\subsection{Motivations and Contributions}
Given the discussion above, one major challenge associated with PELB-driven mmWave MIMO positioning is the difficulty in obtaining accurate and timely bidirectional CSI.
Another challenge is how to exploit the additional degree of freedom (DoF) brought by wideband systems with multipath and multi-subcarrier signals\footnote{In our systems, $\operatorname{DoF} \propto \min\{N_\mathrm{t}, N_\mathrm{r}\} \times N_\mathrm{eff} $, where $ N_\mathrm{t} $ and $ N_\mathrm{r} $ are the numbers of transmit and receive antennas, respectively, and $N_{\mathrm{eff}} = \min\{ N_\mathrm{c}, \frac{B}{B_\mathrm{c}} \}$ denotes is the number of independent subcarrier clusters~\cite{su2025joint,tse2005fundamentals,wang2020optimal}, $B_\mathrm{c}$ is the coherence bandwidth.}.

Motivated by this, we aim to accurately characterize the PELB for multi-path wideband mmWave MIMO systems, and then optimize the signal configuration to guide and improve positioning accuracy. 
Different from the previous PELB studies in~\cite{10370743, 9011751}, this paper proposes to derive the PELB by efficiently fusing multi-dimensional information, including multi-path, multi-subcarrier, and multi-user signals. Multi-dimensional information fusion provides additional system DoF, thereby enhancing positioning accuracy. 
In addition, we specifically extend the proposed mmWave positioning scheme to accommodate more challenging mobile scenarios. In this case, base stations (BSs) need to adaptively adjust beamformers to track variations in the user's position, thereby ensuring the optimal signal configuration for mmWave positioning. To this end, we develop a ping-pong positioning framework to tackle the complicated position estimation and tracking tasks in wideband mmWave MIMO-OFDM systems.

The main contributions of this paper are as follows:
\begin{itemize}
\item A novel device-based\footnote{In device-free radar systems, only echo signals are captured, limiting sensing to the one-way channel from the target to the radar, which compromises positioning precision. In contrast, device-based methods enable the exchange of bidirectional CSI, offering richer channel information to enhance positioning accuracy.} PELB-driven ping-pong positioning framework is designed, in which the BS and user equipment (UE) alternately send adaptive beamforming signals for position estimation. The framework leverages pilots actively transmitted by the UE to enable a bidirectional channel probing mechanism, allowing probing of both the UE-to-BS and BS-to-UE channels. All beamformers are adaptively scheduled based on the locally evaluated PELB. In the first ping-pong round, the BS receives omnidirectional pilots from a UE to perform an uninformed estimation of channel parameters and the user's position. Subsequently, the BS designs beamforming signals to accommodate the PELB. Then, the UE iteratively refines the estimates according to the residuals generated from a reconstructed signal using the initial estimates. From the second round onward, both the BS and the UE utilize the previous round's estimates to iteratively refine the channel and position estimates. 
This uninformed-to-informed approach does not require any prior knowledge of the CSI or user positions, while effectively handling channel dynamics in the context of PELB optimization.

\item By leveraging multi-subcarrier and multipath information to provide additional DoF to improve position accuracy and robustness, a multi-dimensional information fusion-assisted positioning method is proposed. The multi-dimensional information includes multi-subcarriers, multi-path, and multi-user signal. Specifically, we first propose a multi-subcarrier collaborative positioning error lower bound (MSCPEB) to evaluate position accuracy and guide beamformer design. We show that multi-subcarrier signals follow a multivariate Gaussian distribution and further prove that the MSCPEB can achieve a lower PELB than that achieved by the arithmetic mean of the PELBs of the individual subcarriers.
Moreover, we develop a multipath collaborative position method that exploits multipath effects. This method employs path reliability quantification to design weighted cost functions for further improving positioning accuracy.

\item To optimize wideband mmWave positioning accuracy, we formulate an MSCPEB minimization problem for hybrid beamformer and combiner design. Initially, we introduce a subcarrier correlation factor to quantify the contribution of individual subcarriers to the MSCPEB, which effectively decouples the beamformers associated with each subcarrier for simplified MSCPEB optimization. An efficient alternating optimization (AO) algorithm is designed to solve the resulting problem. 
Furthermore, by Schur complement properties, semidefinite relaxation (SDR), and matrix decomposition, we convexify this non-convex MSCPEB minimization problem, and derive closed-form solutions in the AO algorithm. 
\end{itemize}

\subsection{Organization and Notation}
The structure of this paper is organized as follows. Section~\ref{sec:model} describes the system model, constructs the subcarrier correlation factor, and proposes the ping-pong positioning framework. In Section~\ref{sec:CRLB}, we derive and analyze the MSCPEB. In Section~\ref{sec:optimization}, we formulate an MSCPEB minimization problem for hybrid beamformer and combiner design and present an AO algorithm. Section~\ref{sec:estimation} discusses an uninformed-to-informed channel parameter estimation method and a multipath collaborative positioning method. Section~\ref{sec:ana} discusses the computational complexity and convergence behavior of the proposed method. The effectiveness of the proposed method is validated through simulations in Section~\ref{sec:simu}, followed by conclusions drawn in Section~\ref{sec:conc}.

\emph{Notations:} Boldface lowercase and uppercase letters denote vectors and matrices, respectively. $(\cdot)^\mathsf{T}$, $(\cdot)^\mathsf{H}$, $(\cdot)^{-1}$, $(\cdot)^\dagger$ denote the operation of transpose, conjugate transpose, inverse, pseudo-inverse of a vector or a matrix, respectively.  
$\Vert\cdot\Vert_2$ and $\Vert\cdot\Vert_\mathsf{F}$ are L2-norm and F-norm. 
$f_x(x;a)$ denotes the probability distribution function (PDF) of $x$ parameterized by $a$.
$f_{x|y}(x|y;a)$ denotes the PDF of $x$ conditional on $y$ parameterized by $a$.
$\mathcal{CN}(\cdot)$ represents the complex normal distribution. 
$f_\mathcal{CN}(\cdot)$ denotes the complex Gaussian PDF.
$\mathbb{E}_x\{\cdot|a\}$ denotes the expectation for $x$ conditional on $a$. 
$\otimes$ denotes the Kronecker product. 
$\Re\{\cdot\}$ denotes the real part of a complex number.
$\mathrm{rank}\{\cdot\}$ and $\mathrm{tr}\{\cdot\}$ denote the rank and trace of a matrix. 
The $(\imath, \jmath)$-th entry of the matrix $\mathbf{X}$ is denoted by $[\mathbf{X}]_{\imath, \jmath}$. 
$\mathrm{arg}(\cdot)$ and $|\cdot|$ mean taking the phase angle and the modulus, respectively. $\mathbf{I}_N$ is an $N \times N$ identity matrix. 
$\operatorname{arctan2}(y, x)$ returns the four-quadrant inverse tangent of $y$ and $x$.
$A \propto B$ means A is proportional to B.

\section{System Model}\label{sec:model}
This section describes a multi-user wideband mmWave MIMO-OFDM positioning system in which a BS serves $K$ users. We first present the geometry model, channel model, and signal model. Then, we provide an overview of the proposed ping-pong positioning framework.

\subsection{Geometry Model} \label{subsec:geometry}
Assume that there are $L_k$ paths between the BS and the $k$-th UE. For the $\ell_k$-th path, $\alpha_{\ell_k}$, $\tau_{\ell_k}$, $\theta_{\mathrm{t},\ell_k}$, $\phi_{\mathrm{t},\ell_k}$, $\theta_{\mathrm{r},\ell_k}$, and $\phi_{\mathrm{r},\ell_k}$ represent the complex channel gain, the time of arrival (ToA) from the first transmit antenna to the first receive antenna, the elevation and azimuth angles of departure (AoD), and the elevation and azimuth angles of arrival (AoA), respectively. Under the far-field planar wavefront assumption, AoA and AoD from all transmit antennas to all receive antennas are identical. Based on this, the geometric relationships between the BS, the $k$-th UE, and the scattering point (SP) passed through the $\ell_k$-th path are expressed as follows:
\begin{align}
\theta_{\mathrm{t},\ell_k} &=\operatorname{arccos}\big((z_{\mathrm{s},\ell_k}-z_{\mathrm{t}})/\|\mathbf{p}_{\mathrm{s},\ell_k}-\mathbf{p}_{\mathrm{t}}\|_2\big),
\label{geometric2}\\
\phi_{\mathrm{t},\ell_k}&=\operatorname{arctan2}\left(y_{\mathrm{s},\ell_k}-y_{\mathrm{t}},x_{\mathrm{s},\ell_k}-x_{\mathrm{t}}\right), 
\label{geometric1}\\
\theta_{\mathrm{r},\ell_k}&=\operatorname{arccos}\big((z_{\mathrm{s},\ell_k}-z_{\mathrm{r},k})/\|\mathbf{p}_{\mathrm{s},\ell_k}-\mathbf{p}_{\mathrm{r},k}\|_2\big),
\label{geometric4}\\
\phi_{\mathrm{r},\ell_k}&=\operatorname{arctan2}\left(y_{\mathrm{s},\ell_k}-y_{\mathrm{r},k},x_{\mathrm{s},\ell_k}-x_{\mathrm{r},k}\right),
\label{geometric3}\\
\tau_{\ell_k}&=\|\mathbf{p}_{\mathrm{s},\ell_k}-\mathbf{p}_{\mathrm{t}}\|_2/c+\|\mathbf{p}_{\mathrm{s},\ell_k}-\mathbf{p}_{\mathrm{r},k}\|_2/c,
\label{geometric5}
\end{align}
where $c$ is the propagation speed of electromagnetic waves, $\mathbf{p}_{\mathrm{t}}=[x_{\mathrm{t}},y_{\mathrm{t}},z_{\mathrm{t}}]^{\intercal}$, $\mathbf{p}_{\mathrm{r},k}=[x_{\mathrm{r},k},y_{\mathrm{r},k},z_{\mathrm{r},k}]^{\intercal}$ and $\mathbf{p}_{\mathrm{s},\ell_k}=[x_{\mathrm{s},\ell_k},y_{\mathrm{s},\ell_k},z_{\mathrm{s},\ell_k}]^{\intercal}$ are the positions of the BS, the $k$-th UE and the SP passed through the $\ell_k$-th path, respectively.

\subsection{Channel and Received Signal Model} \label{subsec:channel and signal}
To enable 3D localization, we consider a scenario where both the BS and UEs are equipped with uniform planar arrays (UPAs). The BS is equipped with $N_\mathrm{t}=N_\mathrm{t,v}\times N_\mathrm{t,h}$ antennas, where $N_\mathrm{t,v}$ and $N_\mathrm{t,h}$ are the vertical and horizontal dimensions of the transmit array, respectively. Each UE is equipped with  $N_\mathrm{r}=N_\mathrm{r,v}\times N_\mathrm{r,h}$ antennas, where $N_\mathrm{r,v}$ and $N_\mathrm{r,h}$ are the vertical and horizontal dimensions of the receive array, respectively. The spatial-wideband effect in the time domain induces a beam squint effect in the frequency domain. 
Taking this into account, the $n$-th subcarrier is located at a frequency of $f_{\mathrm{c}} + f_{n}$, where $f_\mathrm{c}$ is the carrier frequency, $f_n = (n-1)\Delta f$, $\Delta f =W_\mathrm{c}/N_\mathrm{c}$ is the subcarrier spacing, $W_\mathrm{c}$ is the system bandwidth, $N_\mathrm{c}$ is the number of subcarriers. The channel between the $(n_\mathrm{t,h},n_\mathrm{t,v})$-th transmit antenna and the $(n_\mathrm{r,h},n_\mathrm{r,v})$-th receive antenna at the $n$-th subcarrier is given by \eqref{eq:channeln}, where $\vartheta_{\mathrm{t},\ell_k}=d\cos(\theta_{\mathrm{t},\ell_k})/\lambda_{c}$ and $\varphi_{\mathrm{t},\ell_k}=d\sin(\theta_{\mathrm{t},\ell_k})\cos(\phi_{\mathrm{t},\ell_k})/\lambda_{c}$ are the normalized elevation and the azimuth of AoD related to the $\ell_k$-th path, respectively. $\vartheta_{\mathrm{r},\ell_k}=d\cos(\theta_{\mathrm{r},\ell_k})/\lambda_{c}$ and $\varphi_{\mathrm{r},\ell_k}=d\sin(\theta_{\mathrm{r},\ell_k})\cos(\phi_{\mathrm{r},\ell_k})/\lambda_{c}$ are the normalized elevation and the azimuth of AoA related to the $\ell_k$-th path, respectively. $d$ is the antenna spacing, and $\lambda_\mathrm{c}$ is the wavelength. 

\begin{figure*}[!t]  
\begin{equation}
\begin{aligned}
H_{n_\mathrm{t,v},n_\mathrm{t,h},n_\mathrm{r,v},n_\mathrm{r,h}} (n)=\sum_{\ell_k=1}^{L_k}\alpha_{\ell_k}e^{-j2\pi(f_{c}+f_{n})\tau_{\ell_k}}e^{-j2\pi\left(1+\frac{f_{n}}{f_{c}}\right)\left((n_\mathrm{t,v}-1)\vartheta_{\mathrm{t},\ell_k}+(n_\mathrm{t,h}-1)\varphi_{\mathrm{t},\ell_k}+(n_\mathrm{r,v}-1)\vartheta_{\mathrm{r},\ell_k}+(n_\mathrm{r,h}-1)\varphi_{\mathrm{r},\ell_k}\right)}.
\end{aligned}
\label{eq:channeln}
\end{equation}
\rule{\textwidth}{0.6pt}
\end{figure*}

By defining the array steering vectors as
 \begin{align}
&\bm{a}_{\mathrm{t},n}\left(\vartheta_{\ell_k},\varphi_{\ell_k}\right) =  \frac{1}{\sqrt{N_\mathrm{t}}} \left[1, e^{-j2\pi\left(1+\frac{f_{n}}{f_{\mathrm{c}}}\right)\left(\vartheta_{\mathrm{t},\ell_k}+\varphi_{\mathrm{t},\ell_k}\right)},  \right. \nonumber\\
& \left. \ldots, e^{-j2\pi\left(1+\frac{f_{n}}{f_{\mathrm{c}}}\right)\left((N_{\mathrm{t},v}-1)\vartheta_{\mathrm{t},\ell_k}+(N_{\mathrm{t},h}-1)\varphi_{\mathrm{t},\ell_k}\right)} \right]^\mathsf{T}, \label{eq:arrayres1}\\
&\bm{a}_{\mathrm{r},n}\left(\vartheta_{\ell_k},\varphi_{\ell_k}\right)= \frac{1}{\sqrt{N_\mathrm{r}}} \left[1, e^{-j2\pi\left(1+\frac{f_{n}}{f_{\mathrm{c}}}\right)\left(\vartheta_{\mathrm{r},\ell_k}+\varphi_{\mathrm{r},\ell_k}\right)},  \right. \nonumber\\
& \left. \ldots, e^{-j2\pi\left(1+\frac{f_{n}}{f_{\mathrm{c}}}\right)\left((N_{\mathrm{r},v}-1)\vartheta_{\mathrm{r},\ell_k}+(N_{\mathrm{r},h}-1)\varphi_{\mathrm{r},\ell_k}\right)} \right]^\mathsf{T},\label{eq:arrayres2}
\end{align}
the downlink channel matrix can be structured as
\begin{align}
\mathbf{H}_{k}[n] = & \sqrt{\frac{N_\mathrm{r}N_\mathrm{t}}{L_k}} \sum_{\ell_k=1}^{L_k} \alpha_{\ell_k} e^{-j2\pi(f_\mathrm{c}+f_n) \tau_{\ell_k}} \nonumber\\
& \times  \bm{a}_{\mathrm{r},n}\left(\vartheta_{\ell_k},\varphi_{\ell_k}\right) \bm{a}_{\mathrm{t},n}^\mathsf{H}\left(\vartheta_{\ell_k},\varphi_{\ell_k}\right),
\label{eq:channelmatrix}
\end{align}
{where the channel $\mathbf{H}_{k}[n]\in\mathbb{C}^{N_{\mathrm{r}}\times N_{\mathrm{t}}}$ exhibits reciprocity.
Upon examining \eqref{eq:channelmatrix}, we observe that the subcarrier frequency spacing reflects the information features in the frequency offset and the array phase difference. Specifically, $e^{-j2\pi f_n \tau_{\ell_k}}$ describes the frequency offset caused by the propagation delay $\tau_{\ell_k}$, while $e^{-j2\pi\left(\frac{f_{n}}{f_{\mathrm{c}}}\right)}$ characterizes the array phase difference induced by the carrier frequency offset $f_n$.
To utilize the additional information features carried by multi-subcarrier signals, we extract these two responses to construct the subcarrier correlation factor, as given by
\begin{subequations}
\begin{align} \rho_n=&e^{-j2\pi f_n \tau_{\ell_k}} \times e^{-j2\pi\left(\frac{f_{n}}{f_{\mathrm{c}}}\right)}\left((e^{-j2\pi\left(\frac{f_{n}}{f_{\mathrm{c}}}\right)}\right)^\mathsf{H}  \label{eq:correlation1}\\
=&e^{-j2\pi (n-1)\Delta f \tau_{\ell_k}} \times e^{-j2\pi\left(\frac{(n-1)\Delta f}{f_{\mathrm{c}}}\right)} \nonumber \\
&\times \left((e^{-j2\pi\left(\frac{(n-1)\Delta f}{f_{\mathrm{c}}}\right)}\right)^\mathsf{H}, \label{eq:correlation2}
\end{align}
\label{eq:correlation}
\end{subequations}\\
\noindent where \eqref{eq:correlation2} reveals the relationship between the subcarrier correlation factor and the individual subcarrier index. This is conducive to the universal representation of the mean and covariance of all subcarriers.

\begin{figure*}[h]
  \captionsetup{font=footnotesize}
 \centering
  \includegraphics[width=6.0in]{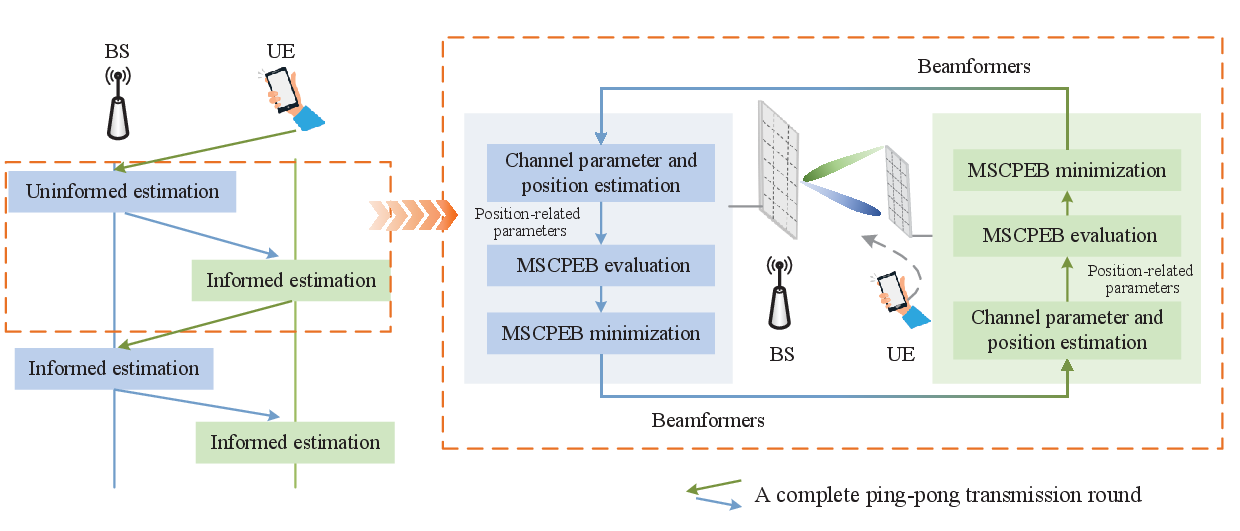}
  \caption{Illustration of the ping-pong positioning framework: The framework comprises three modules: channel parameter and position estimation, MSCPEB evaluation, and MSCPEB minimization. The BS and UE alternately transmit beamforming signals for channel parameter estimation and positioning. The BS performs uninformed estimation in the first round, followed by informed estimation in the subsequent rounds. The estimates of each round serve as the initial conditions for the next informed estimation.}
  \label{framework2}
\end{figure*}

We consider the mutual transmission of $N_\mathrm{s}$ OFDM pilot symbols between the BS and UE. In this scenario, the BS and each UE respectively install $N_\mathrm{t}^{\mathrm{RF}}$ radio frequency (RF) chains and $N_\mathrm{r}^{\mathrm{RF}}$ RF
chains,  where $KN_\mathrm{s} \leq N_\mathrm{t}^{\mathrm{RF}} \ll N_\mathrm{t}$ and $N_\mathrm{s} \leq N_\mathrm{r}^{\mathrm{RF}} \ll N_\mathrm{r}$ need to be satisfied. 
The received signal for the $k$-th user on the $n$-th subcarrier is expressed as
\begin{align}
\mathbf{Y}_{k}[n] &= \mathbf{W}_{\mathrm{BB},k}^\mathsf{H}[n]\mathbf{W}_{\mathrm{RF},k}^\mathsf{H}\mathbf{H}_{k}[n]\mathbf{F}_{\mathrm{RF}}\mathbf{F}_{\mathrm{BB}}[n]\mathbf{s}[n] \nonumber\\
&+ \mathbf{W}_{\mathrm{BB},k}^\mathsf{H}[n]\mathbf{W}_{\mathrm{RF},k}^\mathsf{H}\mathbf{u}_{k}[n] \nonumber\\
&=\mathbf{W}_{\mathrm{BB},k}^\mathsf{H}[n]\mathbf{W}_{\mathrm{RF},k}^\mathsf{H}\mathbf{H}_{k}[n]\mathbf{F}_{\mathrm{RF}}\mathbf{F}_{\mathrm{BB},k}[n]\mathbf{s}_k[n] \nonumber\\
&+\sum_{i \neq k} \mathbf{W}_{\mathrm{BB},k}^\mathsf{H}[n]\mathbf{W}_{\mathrm{RF},k}^\mathsf{H}\mathbf{H}_{k}[n]\mathbf{F}_{\mathrm{RF}}\mathbf{F}_{\mathrm{BB},i}[n]\mathbf{s}_{i}[n] \nonumber\\
&+ \mathbf{W}_{\mathbf{BB},k}^\mathsf{H}[n]\mathbf{W}_{\mathbf{RF},k}^\mathsf{H}\mathbf{u}_{k}[n],
\label{eq:receivedsignals}
\end{align}
where $\mathbf{s}[n]=[\mathbf{s}_1^\mathsf{T}[n],\ldots,\mathbf{s}_K^\mathsf{T}[n]]^\mathsf{T} \in \mathbb{C}^{KN_{s}\times1}$ denotes the symbol vector transmitted by the BS on the $n$-th subcarrier to
UEs, $\mathbf{u}_{k}[n] \in \mathbb{C}^{N_\mathrm{r}\times 1}$ denotes the additive white Gaussian
noise (AWGN) obeying $\mathcal{CN}(0,\sigma_\mathrm{u}^2\mathbf{I}_{N_\mathrm{r}})$, $\sigma_\mathrm{u}^2$ is the noise variance. $\mathbf{F}_{\mathrm{BB}}[n]= [\mathbf{F}_{\mathrm{BB},1}[n],\dots,\mathbf{F}_{\mathrm{BB},K}[n]] \in \mathbb{C}^{N_\mathrm{t}^{\mathrm{RF}}\times KN_\mathrm{s}}$ is the digital precoder on the $n$-th subcarrier.  $\mathbf{F}_{\mathrm{RF}} \in \mathbb{C}^{N_\mathrm{t}\times N_\mathrm{t}^{\mathrm{RF}}}$ is the analog precoder, which adheres to a constant modulus constraint$\big|[\mathbf{F}_{\mathrm{RF}}]_{\imath,\jmath}\big|=1, \forall \imath,\jmath$. The analog precoder uses phase control to adjust the beam's directionality and reduces the complexity of the positioning system.  
$\mathbf{W}_{\mathrm{BB},k}[n]\in\mathbb{C}^{N_\mathrm{r}^{\mathrm{RF}}\times N_\mathrm{s}}$ is the digital combiner of the $n$-th subcarrier for the $k$-th UE, $\mathbf{W}_{\mathrm{RF},k} \in \mathbb{C}^{N_\mathrm{r}\times N_\mathrm{r}^{\mathrm{RF}}}$ is the analog combiner for the $k$-th UE, which also satisfies the constant modulus constraint $\big|[\mathbf{W}_{\mathrm{RF},k}]_{\mathring{\imath},\mathring{\jmath}}\big|=1, \quad \forall k,\mathring{\imath},\mathring{\jmath}$. Without loss of generality, we consider the fully-connected beamforming structure\footnote{For partially-connected structures, the digital precoder and combiner can be modified to block matrices, with each block maintaining the unit-modulus constraint for its entries.} for the derivation of the PELB. The third term in \eqref{eq:receivedsignals} is the signal mean $\bm{\mu}_k[n]$. The sum of the last two terms of \eqref{eq:receivedsignals} represents the interference plus noise of the received signal, and its covariance $\mathbf{C}_{k}[n] \in \mathbb{C}^{N_\mathrm{s}\times N_\mathrm{s}}$ is calculated by
\begin{align}
\mathbf{C}_{k}[n] =& \mathbf{W}^\mathsf{H}_{k}[n]\mathbf{H}_{k}[n]\Big(\sum_{k\neq i}\mathbf{F}_{i}[n]\mathbf{F}^\mathsf{H}_{i}[n]\Big)\mathbf{H}^\mathsf{H}_{k}[n]\mathbf{W}_{k}[n]\nonumber \\
\quad & + \sigma_\mathrm{u}^2\mathbf{W}^\mathsf{H}_{k}[n]\mathbf{W}_{k}[n],
\label{eq:C}
\end{align}

\noindent where {\small $\mathbf{W}_{k}[n]=\mathbf{W}_{\mathrm{RF},k}\mathbf{W}_{\mathrm{BB},k}[n]$} and {\small $\mathbf{F}_{k}[n]=\mathbf{F}_{\mathrm{RF}}\mathbf{F}_{\mathrm{BB},k}[n]$}. 

\subsection{Proposed Ping-Pong Positioning Framework}\label{subsec:framework}
We introduce a novel ping-pong positioning framework, as shown in Fig. \ref{framework2}, which enables channel parameter estimation and positioning through the alternating transmission and reception of beamforming signals between the BS and the UE. All beamformers in this framework are adaptively designed based on endogenously sensed channel parameters and UE positions. Unlike traditional positioning techniques that focus solely on estimation algorithms themselves, our proposed method emphasizes the signal configuration aspects that enhance positioning performance. This framework is designed for slow-mobility scenarios in urban environments.

To be specific, in the first round, the UE transmits omnidirectional pilot signals as the ``serve'' phase, and the BS activates the ping-pong positioning and tracking system via the uninformed estimation. It is worth noting that there is no prior position or CSI available.
In each subsequent ping-pong round, each UE sends adaptive beamforming signals to the BS for informed channel parameter estimation and positioning. Based on the estimated positions, the BS establishes the PELB. Subsequently, the BS minimizes the PELB to guide the hybrid beamforming design. The resulting beamforming signals are transmitted to UEs for informed channel parameter estimation and positioning.
After the UE has estimated the position-related parameters, the associated PELB is established accordingly. Likewise, the hybrid beamforming design at the UE is optimized by minimizing the PELB, thereby initiating the next ping-pong round. The ping-pong rounds are performed for real-time tracking. As long as the user continues to move, the ping-pong process will continue, and the user's position $\mathbf{p}_{\mathrm{r},k}$, the PELB, and the beamformers $\{\mathbf{F}_{\mathrm{BB}}[n],\mathbf{F}_{\mathrm{RF}},\mathbf{W}_{\mathrm{BB},k}[n],\mathbf{W}_{\mathrm{RF},k}\}$ are continuously updated throughout the ping-pong procedure to ensure accuracy.

For ping-pong positioning and tracking systems:
\begin{enumerate}
\item The evaluation and optimization of the MSCPEB are carried out repeatedly for each round, as shown on the right of Fig.~\ref{framework2}. Without loss of generality, we take one transmission round as an example to elaborate on the proposed positioning method. 
\item The optimization methods for the lower bounds of uplink and downlink positioning errors within the same transmission round are similar. For ease of illustration, we take the downlink optimization as an example to present the evaluation and optimization of MSCPEB.
\end{enumerate}

\section{Multi-subcarrier Collaborative Positioning Error Lower Bound} \label{sec:CRLB}

This section introduces an MSCPEB criterion to provide a novel theoretical performance bound for the positioning accuracy of multi-subcarrier systems. In conventional mmWave positioning systems, single-subcarrier-based methods often suffer from limited resolution due to narrow bandwidth and lack of frequency diversity. OFDM-based mmWave systems inherently offer multiple subcarriers, which provide not only more signal snapshots but also additional diversity in delay and angular~\cite{10702556}. To improve the positioning accuracy by leveraging the frequency diversity, we present a concept of MSCPEB through enhanced spatial frequency sampling. To illustrate, consider a radio receiver tuning into multiple frequency channels to locate a transmitter. Each channel provides a unique signal perspective, much like listening to different notes in a musical chord to identify its source. Similarly, by combining frequency-specific ``snapshots'', MSCPEB integrates diverse signal information to achieve a more precise position estimate.

To characterize the PELB from the structure of multi-subcarrier signals, we begin the analysis of MSCPEB with the following reconstruction of received signals across all subcarriers. 
Define the matrix $\mathbf{Y}_k^{\prime}[N_\mathrm{c}]$ as the concatenation of the received signals at the $k$-th user on subcarriers $N_\mathrm{c}$, $N_\mathrm{c}-1, ..., 1$, as given by
\begin{equation}
\mathbf{Y}_k^{\prime}[N_\mathrm{c}]\triangleq\begin{bmatrix}\mathbf{Y}_k^\mathsf{T}[N_\mathrm{c}],\mathbf{Y}_k^\mathsf{T}[N_\mathrm{c}-1],\ldots,\mathbf{Y}_k^\mathsf{T}[1]\end{bmatrix}^\mathsf{T}.
\end{equation}
\\
\noindent Inspired by the analysis of PELB for single-subcarrier systems, we first identify the distribution of the concatenation of the received signals on numerous subcarriers $\mathbf{Y}_k^{\prime}[N_\mathrm{c}]$.

For notational simplicity, let $\ell_k = 1$ represent the line-of-sight (LoS) path. The complete set of channel parameters (including all paths) that need to be estimated is defined as
\begin{align}\label{eq:hlk}
\mathcal{H}_{k} \triangleq &
\{ \alpha_{\ell_k},\tau_{\ell_k},\theta_{\mathrm{t},\ell_k},\phi_{\mathrm{t},\ell_k},\theta_{\mathrm{r},\ell_k}, \nonumber\\
&\quad \phi_{\mathrm{r},\ell_k}:\ell_k=1,2,\ldots,L_k\}.
\end{align}
The subset of channel parameters corresponding to the LoS path is defined as
\begin{equation}
\tilde{\mathcal{H}}_{k}\triangleq\{\alpha_{1},\tau_{1},\theta_{\mathrm{t},1},\phi_{\mathrm{t},1},\theta_{\mathrm{r},1},\phi_{\mathrm{r},1}\}.
\end{equation}
The subset of channel parameters pertaining to the NLoS paths is defined as
\begin{align}
\mathring{\mathcal{H}}_{k}  &\triangleq \mathcal{H}_{k} \backslash \tilde{\mathcal{H}}_{k} 
= \{ \alpha_{\ell_k}, \tau_{\ell_k},\theta_{\mathrm{t},\ell_k}, \phi_{\mathrm{t},\ell_k}, 
 \nonumber\\
&\theta_{\mathrm{r},\ell_k}, \phi_{\mathrm{r},\ell_k} : \ell_k = 2, \ldots, L_k \}.
\end{align}

\begin{lemma}\label{lemma:PDF}
Given $\mathcal{H}_{k}$ and $\mathbf{p}_{\mathrm{r},k}$, the concatenated signal, i.e., random matrix $\mathbf{Y}_k^{\prime}[N_\mathrm{c}]$, follows a multivariate Gaussian distribution. 
\end{lemma}

\begin{IEEEproof}
Refer to Appendix~\ref{appendix3}.
\end{IEEEproof}

According to Lemma~\ref{lemma:PDF}, the PDF of $\mathbf{Y}_k^{\prime}[N_\mathrm{c}]$ is calculated by
\begin{align}
&f_{\mathbf{Y}^{\prime}_k[N_\mathrm{c}]|\mathcal{H}_k}\big(\mathbf{Y^{\prime}}_k[N_\mathrm{c}]|\mathcal{H}_k;\mathbf{p}_{\mathrm{r},k}\big)
\nonumber\\
=&f_\mathcal{CN}\big(\mathbf{Y}^{\prime}_k[N_\mathrm{c}];\bm{\mu}^{\prime}_k[N_\mathrm{c}],\mathbf{C}^{\prime}_k[N_\mathrm{c}]\big),
\end{align}\par
\noindent where the mean $\bm{\mu}^{\prime}_k[N_\mathrm{c}]$ and covariance $\mathbf{C}^{\prime}_k[N_\mathrm{c}]$ are derived as
\begin{align}
\bm{\mu}^{\prime}_k[N_\mathrm{c}]&= \bm{\mu}_k[n] \otimes \bm{\rho}^\mathsf{T}, \label{eq:rho1} \\
\mathbf{C}^{\prime}_k[N_\mathrm{c}]&=\operatorname{diag} \left( \mathbf{C}_{k}[n] \otimes \bm{\rho} \right).
\label{eq:rho2}
\end{align}

\begin{remark}
From \eqref{eq:rho1} and \eqref{eq:rho2}, we define $\bm{\rho}= \begin{bmatrix} \rho_1 & \rho_2 & \cdots & \rho_{N_\mathrm{c}} \end{bmatrix}$ as a subcarrier correlation factor vector, where $\rho_n$ is constructed by \eqref{eq:correlation}. The subcarrier correlation factor matrix encapsulates the inter-subcarrier correlations into a vectorized form, which provides a unified representation of the means and covariance for all subcarriers. This not only simplifies the computation of the MSCPEB contributed by multi-subcarrier cooperation for positioning, but also eliminates the coupling of multi-subcarrier beamformers within the MSCPEB, thus facilitating the formulation of an optimization problem of MSCPEB minimization.
\end{remark}

The mean square error (MSE) of the estimator $\hat{\mathbf{p}}_{\mathrm{r},k}$ is used as a performance metric. The MSE of $\hat{\mathbf{p}}_{\mathrm{r},k}$ with respect to the observation signal $\mathbf{Y}_k^\prime[N_\mathrm{c}]$, considering the parameters in the union set $\mathcal{H}_k$ as nuisance parameters, can be lower bounded by

\begin{subequations}
\begin{align}
& \mathbb{E}_{\mathbf{p}_{\mathrm{r},k}} \Big\{\mathbb{E}_{\mathbf{Y}_k^\prime[N_\mathrm{c}]}\{\|\hat{\mathbf{p}}_{\mathrm{r},k}-\mathbf{p}_{\mathrm{r},k}\|^{2}\}\Big\}\\
=&\mathbb{E}_{\mathbf{p}_{\mathrm{r},k}}\Big\{ \mathbb{E}_{\mathcal{H}_{k}}\Big\{\mathbb{E}_{\mathbf{Y}_k^\prime[N_\mathrm{c}]}|\mathcal{H}_{k}\Big\{\|\hat{\mathbf{p}}_{\mathrm{r},k}-\mathbf{p}_{\mathrm{r},k}\|^{2}|\mathcal{H}_{k}\Big\}\Big|\mathbf{p}_{\mathrm{r},k}\Big\}\Big\} \label{eq:SPEB1}\\
\geqslant& \mathbb{E}_{\mathbf{p}_{\mathrm{r},k}}\Big\{ \mathbb{E}_{\mathcal{H}_{k}}\Big\{\mathrm{tr}\Big(\mathbf{J}^{-1}(\mathbf{p}_{\mathrm{r},k}|\mathcal{H}_{k})\Big)\Big\}\Big|\mathbf{p}_{\mathrm{r},k}\Big\}\label{eq:SPEB2}\\
=&: e(\mathbf{p}).\label{eq:SPEB3}
\end{align}\label{eq:SPEB}
\end{subequations}\\
\noindent Here, \eqref{eq:SPEB1} provides the method for calculating MSE for position estimation. In \eqref{eq:SPEB2}, $\mathbf{J}\big(\mathbf{p}_{\mathrm{r},k}\big|\mathcal{H}_{k}\big)$ represents the FIM concerning positional information, obtained from the observation of $\mathbf{Y}_k^\prime[N_\mathrm{c}]$ for given $\mathcal{H}_{k}$, and the trace of its inverse corresponds to the CRLB. In \eqref{eq:SPEB3}, we define the CRLB for position estimation as the MSCPEB $e(\mathbf{p})$. Table~\ref{tab:MSCPEB} summarizes the traditional and the proposed multi-subcarrier positioning error bounds, including their notations and definitions.

\begin{table}[t]
\centering 
\renewcommand{\arraystretch}{1.8}
\caption{Summary of Traditional and Multi-Subcarrier Positioning Error Lower Bounds} 
\label{tab:MSCPEB}
\begin{tabular}{p{2.4cm}|p{1cm}|p{3.4cm}}
\hline
\textbf{Bound} & \textbf{Notation} & \textbf{Definition} \\
\hline
Classical CRLB & CRLB & Lower bound on variance of any unbiased estimator  \\
Single-subcarrier PELB & $e_n(p)$ & CRLB for single-subcarrier position estimation \\
Multi-subcarrier collaborative PELB (MSCPEB)& $e(p)$ & CRLB for multi-subcarrier position estimation \\
\hline
\end{tabular}
\end{table}

The random presence of the LoS path is represented using a binary indicator $\delta_k\in\{0,1\}$, which follows a Bernoulli distribution, i.e.,
\begin{equation}
\delta_{k}=
\begin{cases} 
1, & \text{with probability } {p}_{\text{\tiny LoS}}(\tau_{\ell_k}c) \\
0, & \text{with probability } 1-{p}_{\text{\tiny LoS}}(\tau_{\ell_k}c)
\end{cases}
\end{equation}
The LoS probability $p_{\text{\tiny LoS}}(\tau_{\ell_k}c)$ depends on the path distance $\tau_{\ell_k}c$. 
The MSCPEB can be calculated by
\begin{align}\label{eq:ep}
e(\mathbf{p}) &=\mathbb{E}_{\mathbf{p}_{\mathrm{r},k}}\Big\{\mathbb{E}_{\mathcal{H}_{k}}\Big\{\mathbb{E}_{\delta_{k}}\Big\{\mathrm{tr}\Big(\mathbf{J}^{-1}(\mathbf{p}_{\mathrm{r},k}|\mathcal{H}_{k},\delta_{k})\Big)\Big|\mathcal{H}_{k}\Big\}\Big|\mathbf{p}_{\mathrm{r},k}\Big\}\Big\} \nonumber\\
& \stackrel{(a) }{\approx} \frac1K\sum_{k=1}^K \mathrm{tr}\Big(\mathbf{J}^{-1}\big(\mathbf{p}_{\mathrm{r},k}\big|\mathcal{H}_{k}\big)\Big) \nonumber\\
&\stackrel{(b)}{=}\frac1K\sum_{k=1}^K\bigl(p_{\text{\tiny LoS}}(\tau_{\ell_k}c)\bigr)^{\delta_k} \mathrm{tr}\Big(\mathbf{J}^{-1}\big(\mathbf{p}_{\mathrm{r},k}\big|\tilde{\mathcal{H}}_{k}\big)\Big) \nonumber\\
&+ \frac{1}{K} \sum_{k=1}^K \bigl(1-p_{\text{\tiny LoS}}(\tau_{\ell_k}c)\bigr)^{1-\delta_k} \mathrm{tr}\Big(\mathbf{J}^{-1}\big(\mathbf{p}_{\mathrm{r},k}\big|\mathring{\mathcal{H}}_{k}\big)\Big),\end{align}
where (a) involves averaging the MSCPEB over multiple users, and (b) comes from the expectation of the MSCPEB over different propagation scenarios.

$\mathbf{J}\big(\mathbf{p}_{\mathrm{r},k}\big|\mathcal{H}_{k}\big)$ can be transformed using the chain rule, i.e.,
\begin{align}\label{eq:chain}
\mathbf{J}\big(\mathbf{p}_{\mathrm{r},k}\big|\mathcal{H}_{k}\big)=&\Big(\frac{\partial\eta_{\ell_k}^\mathsf{T}}{\partial \mathbf{p}_{\mathrm{r},k}}\otimes\mathbf{1}_{N_\mathrm{c}\times N_\mathrm{c}N_\mathrm{s}}\Big)\mathbf{J}\big(\eta_{\ell_k}\big|\mathcal{H}_{k}\big) \nonumber\\
&\times \Big(\frac{\partial\eta_{\ell_k}^\mathsf{T}}{\partial \mathbf{p}_{\mathrm{r},k}}\otimes\mathbf{1}_{N_\mathrm{c}\times N_\mathrm{c}N_\mathrm{s}}\Big)^\mathsf{T},
\end{align}
where $\mathbf{1}_{N_\mathrm{c}\times N_\mathrm{c}N_\mathrm{s}} \in \mathbb{C}^{N_\mathrm{c} \times N_\mathrm{c}N_\mathrm{s}}$ represents a vector where all elements are equal to 1. For ease of calculation, $\eta_{\ell_k}\triangleq[\tau_{\ell_k},\theta_{\mathrm{t},\ell_k},\phi_{\mathrm{t},\ell_k},\theta_{\mathrm{r},\ell_k},\phi_{\mathrm{r},\ell_k}]^{\mathsf{T}}$. It should be noted that the chain rule in~\eqref{eq:chain} applies to the transformation of the FIM along the path $\ell_k$. As shown in~\eqref{eq:hlk}, $\mathcal{H}_k$ in the conditional FIM $\mathbf{J}\big(\eta_{\ell_k}\big|\mathcal{H}_{k}\big)$ represents the complete set of channel parameters (including all paths). Therefore, the traversal over all paths is handled in~\eqref{eq:SPEB} through the expectation operator. The transformation
matrix $\frac{\partial\eta_{\ell_k}}{\partial \mathbf{p}_{\mathrm{r},k}}$ is derived in Appendix \ref{Transmatrix}. The FIM $\mathbf{J}\big(\eta_{\ell_k}\big|\mathcal{H}_{k}\big)$ is defined as \eqref{eq:FIM}, which is shown at the top of the next page. More importantly, we prove that \eqref{eq:FIM} does not exceed the arithmetic mean of the PELBs of the individual subcarriers in the following lemma.

\begin{lemma}\label{lemma:J}
For theoretical considerations, the MSCPEB does not exceed the arithmetic mean of the PELBs of the individual subcarriers.
\end{lemma}
\begin{IEEEproof}
Refer to Appendix \ref{appendixJ}.
\end{IEEEproof}

Since $\mathbf{Y}_k^\prime[N_\mathrm{c}]$ and its interference plus noise are Gaussian distributed, the $({\imath},{\jmath})$-th element of the conditional FIM in \eqref{eq:FIM} is expressed as \eqref{eq:J1}, and shown at the top of this page, where $m_{{\imath},{\jmath}}$ represents the variance of the $(\imath,\jmath)$-th element in FIM, $v_{{\imath},{\jmath}}$ represents the mean of the $(\imath,\jmath)$-th element in FIM. The derivatives $\frac{\partial\bm{\mu}_k[n]}{\partial[\eta_{k}]_{{\imath}}}$ is provided in Appendix~\ref{FIMderivatives}. In optimization problems, we usually ignore $m_{{\imath},{\jmath}}$ to simplify calculations~\cite{boyd2004convex}. 

\begin{figure*}[!t]
\begin{align}
\mathbf{J}(\eta_{\ell_k}|\mathcal{H}_{k})  
&\triangleq  \mathbb{E}\Big\{\frac{\partial\ln f_{\mathbf{Y}_k^\prime[N_\mathrm{c}]|\mathcal{H}_{k}}\big(\mathbf{Y}_k^\prime[N_\mathrm{c}]\Big| \mathcal{H}_{k};\eta_{\ell_k}\big)}{\partial\eta_{\ell_k}} 
\cdot  \frac{\partial\ln f_{\mathbf{Y}_k^\prime[N_\mathrm{c}]|\mathcal{H}_{k}}\left(\mathbf{Y}_k^\prime[N_\mathrm{c}]\right|\mathcal{H}_{k};\eta_{\ell_k})}{\partial\eta_{\ell_k}^{\mathsf{T}}}\Big|\mathcal{H}_{k}\Big\}.
\label{eq:FIM}\\
\left[\mathbf{J}(\eta_{\ell_k}|\mathcal{H}_{k}) \right]_{{\imath},{\jmath}} 
&=\underbrace{\mathrm{tr}\Big((\mathbf{C}_k^\prime[N_\mathrm{c}])^{-1}\frac{\partial\mathbf{C}_k^\prime[N_\mathrm{c}]}{\partial[\eta_{\ell_k}]_{{\imath}}} (\mathbf{C}_k^\prime[N_\mathrm{c}])^{-1}\frac{\mathbf{C}_k^\prime[N_\mathrm{c}]}{\partial[\eta_{\ell_k}]_{{\jmath}}}\Big)}_{\triangleq m_{{\imath},{\jmath}}}
+ \underbrace{2\Re\Big\{\frac{\partial\big(\bm{\mu}_k^\prime[N_\mathrm{c}]\big)^{\mathsf{H}}}{\partial[\eta_{\ell_k}]_{{\imath}}}(\mathbf{C}_k^\prime[N_\mathrm{c}])^{-1}\frac{\partial\bm{\mu}_k^\prime[N_\mathrm{c}]}{\partial[\eta_{\ell_k}]_{{\jmath}}}\Big|\mathcal{H}_{k}\Big\}}_{\triangleq v_{{\imath},{\jmath}}}.
\label{eq:J1}
\end{align}
\rule{\textwidth}{0.6pt}
\end{figure*}

\section{Positioning Error Lower Bound Minimization: Proposed Solution} \label{sec:optimization}
This section investigates the problem of minimizing PELB to create a beam configuration conducive to improving positioning performance and efficiency.
Our goal is to find the optimal precoders $\mathbf{F}_{\mathrm{BB}}[n]$ and $\mathbf{F}_{\mathrm{RF}}$ at the BS and the combiners $\mathbf{W}_{\mathrm{RF},k}$ and $\mathbf{W}_{\mathrm{BB},k}[n]$ at the $k$-th user by minimizing the PELB, subject to the transmit power constraint per subcarrier and the constant-modulus constraints for the phase shifts. 
The objective function in \eqref{eq:objective function} is derived from the MSCPEB, and the aim is to minimize this error bound, as it describes the theoretical achievable accuracy of position estimation. 
As a result, the original optimization problem is formulated as: \begin{subequations}\label{P1} 	
\begin{align}	
\textbf{P1}:\quad &\underset{\scalebox{0.6}{$ \mathbf{F_{\mathrm{RF}}}, \mathbf{F}_{\mathrm{BB}}[n], \mathbf{W_{\mathrm{RF},k}}, \mathbf{W}_{\mathrm{BB},k}[n]$}}{\min}\, \quad \frac1K\sum_{k=1}^K \mathrm{tr}\Big(\mathbf{J}^{-1}\big(\mathbf{p}_{k}\big|\mathcal{H}_{k}\big)\Big) \tag{26a} \label{eq:objective function}\\
\text{ s.t.} \quad \; \:   
&\|\mathbf{F}_{\mathrm{RF}}\mathbf{F}_{\mathrm{BB}}[n]\|_\mathsf{F}^{2} \leq P_n, \tag{26b}\label{eq:power1}\\
& \big|[\mathbf{F}_\mathrm{RF}]_{\imath,\jmath}\big|=1, \quad \forall \imath,\jmath, \tag{26c}\label{eq:modulus1}\\
&\big|[\mathbf{W}_{\mathrm{RF},k}]_{\mathring{\imath},\mathring{\jmath}}\big|=1, \quad \forall k,\mathring{\imath},\mathring{\jmath}, \tag{26d}\label{eq:modulus2}
\end{align}
\end{subequations}
where~\eqref{eq:objective function} involves averaging the MSCPEB of multiple users, and a more detailed derivation is shown in~\eqref{eq:ep};~\eqref{eq:power1} accounts for the transmit power budget; \eqref{eq:modulus1} and \eqref{eq:modulus2} represent constant-modulus constraints for the phase shifts of the transceivers \footnote{While the derivation of MSCPEB and beam optimization in this work employs simulated phase shifters with infinite resolution, the method remains applicable to discrete systems with finite resolution. This is achieved by quantizing the continuous solution to the available discrete phase set.}.  We observe that the optimization variables, $\mathbf{F_{\mathrm{RF}}}$ and $\mathbf{F}_{\mathrm{BB}}[n]$ are precoders specifically designed for multi-user scenarios. Consequently, the optimization objective involves averaging over all users. It should be noted that \textbf{P1} is non-convex and NP-hard, due to the complex conditional Fisher information, the highly-coupled beamformers, and the non-convex constant-modulus constraint. Consequently, obtaining the global optimal solution to \textbf{P1} is remarkably arduous. To this end, we develop an AO framework to efficiently solve \textbf{P1}, where the local optimal hybrid precoders and combiners can be guaranteed with satisfactory positioning precision.

Considering that the multi-user interference and noise are influenced by the channel environment and beamformers, the conditional FIM $\mathbf{J}\big(\mathbf{p}_{k}\big|\mathcal{H}_{k}\big)$ no longer retains its original quadratic form. Therefore, the original optimization methods cannot be directly applied here, and \textbf{P1} needs to be effectively convexified.
Firstly, the product of digital and analog precoder is considered as a whole, and our objective is to find the hybrid precoder $\mathbf{F}[n]=\mathbf{F}_\mathrm{RF}\mathbf{F}_\mathrm{BB}[n]$ that minimizes the MSCPEB, which can be expressed as
\begin{subequations}\label{P2} 
\begin{align}	
\!\!\!\!\textbf{P2}:\quad &\underset{\mathbf{F}[n]}{\operatorname*{min}}\quad \frac1K\sum_{k=1}^K \mathrm{tr}\Big(\mathbf{J}^{-1}\big(\mathbf{p}_{\mathrm{r},k}\big|\mathcal{H}_{k}\big)\Big) \\
\text{s.t.} \quad  \:   
& ||\mathbf{F}[n]||_\mathsf{F}^{2}\leq P_n.
\end{align}
\end{subequations}
The power constraint $||\mathbf{F}[n]||_\mathsf{F}^{2} \leq P_n$ and the trace-based constraint $\mathrm{tr}\left\{\mathbf{V}[n]\right\}\leq P_n$ are mathematically equivalent. Thus, we reformulate \textbf{P2} as
\begin{subequations}\label{P3} 
\begin{align}	
\!\!\!\!\textbf{P3}:\quad &\underset{\mathbf{V}[n]}{\min}\, \quad \frac1K\sum_{k=1}^K \mathrm{tr}\Big(\mathbf{J}^{-1}\big(\mathbf{p}_{k}\big|\mathcal{H}_{k}\big)\Big) \\
\text{ s.t.} \quad \; \:   
&\mathrm{tr}\left\{\mathbf{V}[n]\right\}\leq P_n, \label{eq:power2}\\
& \mathbf{V}[n]\succeq0,\label{eq:v}\\
&\mathrm{rank}\left\{\mathbf{V}[n]\right\}\leq K \label{eq:rank1}.
\end{align}
\end{subequations}

However, the conditional FIM $\mathbf{J}\big(\mathbf{p}_{k}\big|\mathcal{H}_{k}\big)$ contains a large number of matrix derivatives, and the corresponding solutions and inverse operations are complicated. To this end, we introduce an auxiliary variable $\bm{\Omega} \in \mathbb{R}^{3N_\mathrm{c} \times 3N_\mathrm{c}}$ to relax the objective function, which satisfies the following conditions
\begin{equation}\label{simplifyFIM} \bm{\Omega}_k \succeq \mathbf{E}^\mathsf{T}\mathbf{J}^{-1}\big(\mathbf{V}[n]\big)\mathbf{E}, \quad \forall k,\end{equation}
where $\mathbf{E}$ is a $3N_\mathrm{c} \times 3N_\mathrm{c}$ identity matrix. To ensure invertibility and the applicability of the CRLB, the FIM is typically assumed to be positive definite, provided that the parameters are identifiable and the observation data is sufficiently comprehensive.
Therefore, \eqref{simplifyFIM} can be  reformulated using the Schur complement of the FIM to simplify computations, as follows:
\begin{equation}\label{eq:schur}\begin{bmatrix}\bm{\Omega}_k&\mathbf{E}^\mathsf{T}\\\mathbf{E}&\mathbf{J}\big(\mathbf{V}[n]\big)\end{bmatrix}\succeq \bm{0},\quad\forall k.\end{equation}
Then, the problem is further transformed into 
\begin{subequations}\label{P4}
\begin{align}	
\!\!\!\!\textbf{P4}:\quad &\underset{\mathbf{V}[n],\bm{\Omega}_k}{\min}\, \quad \frac1K\sum_{k=1}^K\mathrm{tr}\left\{\bm{\Omega}_k\right\} \notag \\
\text{ s.t.} \quad \; \:  
& \begin{bmatrix}\bm{\Omega}_k&\mathbf{E}^\mathsf{T}\\\mathbf{E}&\mathbf{J}\big(\mathbf{V}[n]\big)\end{bmatrix}\succeq \bm{0},\quad\forall k,\\
& (\ref{eq:power2}), (\ref{eq:v}). \notag
\end{align}
\end{subequations}
Problem \textbf{P4} can be solved by applying SDR, which allows us to eliminate the rank constraint \eqref{eq:rank1}. 
Accordingly, the solution to \textbf{P4} can be attained using the semidefinite programming (SDP) solver in MATLAB. 
Even in the absence of the rank constraint, the optimal solution $\mathbf{V}^*[n]$ to \textbf{P4} is guaranteed to be rank-$K$~\cite{boyd2004convex}.

After obtaining $\mathbf{V}[n]$, we can recover the product of the hybrid precoder via the matrix decomposition method~\cite{8804387}. 
Defining $R \triangleq \operatorname{rank}\{\mathbf{V}[n]\} \leq K$, we arrive at $\operatorname{rank}\{\mathbf{F[n]}\}=R \leq K$ from relation $\mathbf{V}[n]\triangleq \mathbf{F}[n]\mathbf{F}^\mathsf{H}[n]$. A solution satisfying $\operatorname{rank}\{\mathbf{F[n]}\}=R \leq K$ can be achieved by expanding $\mathbf{F}[n]$ and $\mathbf{V}[n]$ via compact singular value decomposition (SVD) and eigendecomposition (ED).
The SVD of $\mathbf{F}[n]$ and the ED of $\mathbf{V}[n]$ are given by
\begin{align}
\mathbf{F}[n]&=\mathbf{Z}_\mathrm{left}\bm{\Theta}\mathbf{Z}_\mathrm{right}^\mathsf{H},\label{eq:F}\\
\mathbf{V}[n]&=\mathbf{Q}\bm{\Psi}\mathbf{Q}^\mathsf{H},\label{eq:V}
\end{align}
where $\mathbf{Z}_\mathrm{left}=[\mathbf{z}_{\mathrm{left},1}\cdots\mathbf{z}_{\mathrm{left},R}]\in\mathbb{C}^{N_{\mathrm{t}}\times R}$ collects left singular vectors of $\mathbf{F}[n]$, $\mathbf{Z}_\mathrm{right}=[\mathbf{z}_{\mathrm{right},1}\cdots\mathbf{z}_{\mathrm{right},R}]\in\mathbb{C}^{R\times R}$ collects right singular vectors of $\mathbf{F}[n]$, 
$\bm{\Theta}$ is a diagonal matrix with its diagonal elements being singular values $ \{\zeta_1\cdots\zeta_R\}$,
$\mathbf{Q}=[\mathbf{q}_{1}\cdots\mathbf{q}_{R}]\in\mathbb{C}^{N_{\mathrm{t}}\times R}$ collects eigenvectors of $\mathbf{V}[n]$, and $\bm{\Psi}$ is a diagonal matrix with its diagonal elements being eigenvalues $\{ \lambda_1\cdots\lambda_R\}$.

Substituting the results of~\eqref{eq:F} and~\eqref{eq:V} into $\mathbf{V}[n] = \mathbf{F}[n]\mathbf{F}^\mathsf{H}[n]$, we arrive at the following matrix equation
\begin{equation}\mathbf{Z}_\mathrm{left}\bm{\Theta}^{2}\mathbf{Z}_\mathrm{right}^\mathsf{H}=\mathbf{Q}\bm{\Psi}\mathbf{Q}^\mathsf{H}.\label{eq:ZthetaZ}\end{equation}
The solution to~\eqref{eq:ZthetaZ} involves matching the eigenvectors and eigenvalues on both sides of the equation, i.e.,
\begin{align}\mathbf{z}_{\mathrm{left},r} =\mathbf{q}_r, \;& r=1,\cdots, R.\\
\zeta_r  =\sqrt{\lambda_r},\; & r=1,\cdots, R.\end{align}

Analogously to \textbf{P4}, we recover the hybrid combiner by treating the product of the digital and analog precoders as a single entity. Let $\mathbf{O}_k[n]\triangleq \mathbf{W}_k[n]\mathbf{W}_k^\mathsf{H}[n]$. Our objective is to find the hybrid combiner that minimizes the MSCPEB, i.e.,
\begin{subequations}
\begin{align}	
\!\!\!\!\textbf{P5}:\quad &\underset{\mathbf{O}_k[n],\bm{\Omega}_k}{\min}\, \quad \frac1K\sum_{k=1}^K\mathrm{tr}\left\{\bm{\Omega}_k\right\}  \\
\text{ s.t.} \quad \; \:  
& \begin{bmatrix}\bm{\Omega}_k&\mathbf{E}^\mathsf{T}\\\mathbf{E}&\mathbf{J}\big(\mathbf{O}_k[n]\big)\end{bmatrix}\succeq \bm{0},\quad\forall k.
\end{align}
\end{subequations}
Problem \textbf{P5} can be tackled using the SDP solver and matrix decomposition.

Furthermore, the analog precoder/combiner and digital precoder/combiner can be decoupled under the orthogonal assumption of the digital precoder \cite{7397861}. Taking the precoder as an example, the closed-form expressions of the analog precoder and combiner are
\begin{align}
[\mathbf{F}_\mathrm{RF}]_{\imath,\jmath}&=e^{j\arg\left(\left[\mathbf{F}[n]\mathbf{F}_\mathrm{uni}[n]^\mathsf{H}\right]_{\imath,\jmath}\right)},\label{eq:close F}\\
\mathbf{F}_\mathrm{BB}[n]^\mathsf{H}\mathbf{F}_\mathrm{BB}[n] &\approx \alpha \mathbf{F}_\mathrm{uni}[n]^\mathsf{H}\alpha\mathbf{F}_{\mathrm{uni}}[n],
\label{eq:close B} 
\end{align}
where $\mathbf{F}_\mathrm{uni}[n]=\dot{\mathbf{V}}_{\mathrm{right}}\mathbf{V}_{\mathrm{left}}^\mathsf{H}$ is a unitary matrix with the same dimension as $\mathbf{F}_\mathrm{BB}[n]$. The term $e^{j\arg\left(\left[\mathbf{F}[n]\mathbf{F}_\mathrm{uni}[n]^\mathsf{H}\right]_{\imath,\jmath}\right)}$ is a complex number with unit modulus. Its phase is the same as the phase of $\left[\mathbf{F}[n]\mathbf{F}_\mathrm{uni}[n]^\mathsf{H}\right]_{\imath,\jmath}$, but its modulus is always 1. $\alpha$ is a constant. $\mathbf{V}_\mathrm{left}\mathbf{S}\dot{\mathbf{V}}_\mathrm{right}^\mathsf{H}$ is the SVD of $\mathbf{F}[n]^\mathsf{H}\mathbf{F}_{\mathrm{RF}}$. $\mathbf{V}_\mathrm{left}$ collects left singular vectors, $\mathbf{V}_\mathrm{right}$ collects right singular vectors. $\dot{\mathbf{V}}_\mathrm{right}$ represents the first $N_\mathrm{s}$ columns of $\mathbf{V}_\mathrm{right}$, $\mathbf{S}$ is a diagonal matrix containing the first $N_\mathrm{s}$ non-zero singular value $\{\sigma_1,\ldots,\sigma_{N_\mathrm{s}}\}$.

The digital and analog combiners also have closed-form expressions similar to \eqref{eq:close F} and \eqref{eq:close B}, from which we can obtain the complete optimal solution of \textbf{P1}.
Above all, the developed AO algorithm for hybrid beamforming is summarized in \textbf{Algorithm}~\ref{alg:alter}.

\begin{algorithm}

\caption{Alternating Optimization for Analog and Digital Beamformers}
\label{alg:alter}
\begin{algorithmic}[1] 
        \STATE \textbf{Initialize:} $\mathbf{F}_\mathrm{BB}[n]^{(0)}$, $\mathbf{W}_{\mathrm{BB},k}[n]^{(0)}$, $\bm{\Omega}_k$, $\mathbf{F}[n]^{(0)}$, $\mathbf{F}_\mathrm{RF}^{(0)}$, $\mathbf{W}_k[n]^{(0)}$, $\mathbf{W}_{\mathrm{RF},k}^{(0)}$.
\WHILE{True}
    \STATE Given $\mathbf{W}_k[n]^{(t)}$, update $\mathbf{F}[n]^{(t+1)}$ by solving \textbf{P4};
    \STATE Given $\mathbf{F}[n]^{(t+1)}$, update $\mathbf{W}_k[n]^{(t+1)}$ by solving \textbf{P5}; 
    \STATE Update $t = t + 1$;
    \IF{accuracy threshold reached}
        \STATE Perform
        $\{\mathbf{F}_{\mathrm{BB}}[n], \mathbf{F}_{\mathrm{RF}}\} \gets \{\mathbf{F}^{\mathrm{opt}}[n] \}$ and $\{\mathbf{W}_{\mathrm{BB},k}[n], \mathbf{W}_{\mathrm{RF},k}\} \gets \{\mathbf{W}^{\mathrm{opt}}_k[n] \}$, respectively;
        \STATE Break;
    \ENDIF
\ENDWHILE
\STATE \textbf{Output:} $\mathbf{F}_{\mathrm{BB}}[n]$, $\mathbf{F}_{\mathrm{RF}}$, $\mathbf{W}_{\mathrm{BB},k}[n]$, $\mathbf{W}_{\mathrm{RF},k}$.
\end{algorithmic}
\end{algorithm}

\section{Channel Parameter and Position Estimation}\label{sec:estimation}
This section provides an algorithm for joint channel parameter and UE position estimation, based on the signal configuration that minimizes the MSCPEB. The uninformed-to-informed estimation of involved channel parameters is first described. 

\subsection{Channel Parameter Estimation}
\subsubsection{Uninformed Estimation}\label{Coarse estimation}
In the first ping-pong round, there is no prior information regarding the CSI or position available. With the received omnidirectional pilot signal at the BS, we perform an uninformed estimate of the position-related parameters. Then, we formulate an optimization problem of uninformed estimation which relies on the correlation between the received signal and the candidate signal reconstructed from the parameter estimates, rather than depending on the channel structure. Both the received signal and the reconstructed signal are multi-subcarrier concatenation signals. As discussed in~\cite{wang2024performance}, time delay estimation is rather challenging under far-field conditions. Although single-subcarrier time delay estimation can be addressed by the methods developed in~\cite{affes1998new,feder1988parameter}, these methods require high signal-to-noise ratios (SNR) and extended observation periods. In contrast, multi-subcarrier systems can significantly enhance the accuracy and robustness of time delay estimation through frequency diversity. Based on this, the optimization problem is formulated as
\begin{equation}\small
\begin{aligned}
&\tau_{\ell_k},\theta_{\mathrm{t},\ell_k},\phi_{\mathrm{t},\ell_k},\theta_{\mathrm{r},\ell_k},\phi_{\mathrm{r},\ell_k}\\
&=\underset{\tau\in\mathbb{R}_\tau,\{\theta_{\mathrm{t},\ell_k},\phi_{\mathrm{t},\ell_k}\}\in\mathbb{R}_\mathrm{T},\{\theta_{\mathrm{r},\ell_k},\phi_{\mathrm{r},\ell_k}\}\in\mathbb{R}_\mathrm{R}}{\operatorname{arg\,max}}|\mathrm{vec}(\mathbf{Y}^{\prime}[n])^\mathsf{H}\cdot\mathrm{vec}(\hat{\mathbf{Y}}^{\prime}[n])|,
\end{aligned}
\end{equation}
where $\hat{\mathbf{Y}}[n]$ is the candidate signal reconstructed from the estimated parameters. The ToF, AoA and AoD are searched separately on independent grids. Specifically, $\mathbb{R}_{\tau}$ is a one-dimensional vector,  scretizing the time range $[0,T]$ into $N_{\tau}$ discrete points. $\mathbb{R}_\mathrm{T}$ is a two-dimensional grid, discretizing the three-dimensional (3D) angle of AoA $[0,\pi]\times[0,2\pi]$ into an $N_{\theta}\times N_{\phi}$ grid. $\mathbb{R}_\mathrm{R}$ is a two-dimensional grid, similar to $\mathbb{R}_\mathrm{T}$, discretizing the 3D angle of AoD into an $N_\theta\times N_\phi$ grid.

In the initial uninformed estimation, the path gain $\alpha_{\ell_k}$ is set to 1. The accuracy of $\alpha_{\ell_k}$ directly impacts the quality of received signal reconstruction. To enable subsequent precise signal reconstruction, it is essential to estimate the path gain based on the uninformed estimation results. This problem is described by
\begin{equation}
\label{eq:alpha}\{\alpha_{\ell_k}\}=\arg\min_{\{\alpha_{\ell_k}\}}\;\Lambda_k,
\end{equation}
where $\Lambda_k=\|\mathbf{Y}^{\prime}_k[n]-\hat{\mathbf{Y}}^{\prime}_k[n]]\|_\mathsf{F}^2$ is the residual of the channel parameters estimate. Problem \eqref{eq:alpha} can be rewritten as
\begin{equation}\label{eq:alpha2}\begin{bmatrix}\operatorname{vec}(\hat{\mathbf{Y}^{\prime}}_1[n]),\ldots,\operatorname{vec}(\hat{\mathbf{Y}}^{\prime}_{\ell_k}[n])\end{bmatrix}\begin{bmatrix}\alpha_1\\\vdots\\\alpha_{\ell_k}\end{bmatrix}=\operatorname{vec}(\mathbf{Y}^{\prime}_k[n]),\end{equation}
which can be easily solved using least squares.

\subsubsection{Informed Estimation}\label{Accurate estimation}
After the uninformed estimation, hybrid beamforming is optimized based on the criterion of minimizing MSCPEB.
Then, we perform the informed estimation of channel parameters based on the received beamformed signals. In this case, directly applying classical spatial spectrum estimation may not fully exploit the rich features of the received signal. Therefore, we propose to maximize the signal fitting accuracy to ensure the minimal reconstruction error between the estimated and actual signals.
For each path, informed estimation is expressed by
\begin{equation}
\label{eq:accurate}
\tau_{\ell_k},\theta_{\mathrm{t},\ell_k},\phi_{\mathrm{t},\ell_k},\theta_{\mathrm{r},\ell_k},\phi_{\mathrm{r},\ell_k}=\underset{\tau_{\ell_k},\theta_{\mathrm{t},\ell_k},\phi_{\mathrm{t},\ell_k},\theta_{\mathrm{r},\ell_k},\phi_{\mathrm{r},\ell_k}}{\operatorname{arg\,min}}\Lambda_k.
\end{equation}
To solve \eqref{eq:accurate}, we use the Newton method to update the last round estimation result of each parameter, i.e.,
\begin{equation}
\label{eq:Newton}
\eta_{\ell_k,i}\leftarrow\eta_{\ell_k,i}-\gamma\frac{\partial\Lambda_k}{\partial\eta_{\ell_k,i}}\left(\frac{\partial^2\Lambda_k}{(\partial\eta_{\ell_k,i})^2}\right)^{-1},
\end{equation}
where $\gamma$ is the update step size, $i$ is the index of the channel parameters $\eta_{\ell_k}$. The derivatives in (\ref{eq:Newton}) are presented in Appendix~\ref{sec:Newton}.

\subsection{Multi-Path Collaborative Position Estimation}\label{Position Estimation}

To eliminate the reliance on path resolution and traditional triangular relationships, we transform the position estimation problem from finding traditional triangular relationships to analyzing the relationship between the departure direction vector and the receive direction vector of the signal. We define $\vec{f}_{\mathrm{t,\ell_k}}$ as the transmit direction vector associated with the AoD, and $\vec{f}_{\mathrm{r,\ell_k}}$ as the receive direction vector associated with the AoA. The transmit and receive direction vectors are given by
\begin{subequations}
\begin{align}
\vec{f}_{\mathrm{t,\ell_k}}=\begin{bmatrix}\sin(\hat{\theta}_{\mathrm{t},\ell_k})\cos(\hat{\phi}_{\mathrm{t},\ell_k})\\\sin(\hat{\theta}_{\mathrm{t},\ell_k})\sin(\hat{\phi}_{\mathrm{t},\ell_k})\\\cos(\hat{\theta}_{\mathrm{t},\ell_k})\end{bmatrix},\label{eq:subeq-a}\\
\vec{f}_{\mathrm{r,\ell_k}}=\begin{bmatrix}\sin(\hat{\theta}_{\mathrm{r,\ell_k}})\cos(\hat{\phi}_{\mathrm{r,\ell_k}})\\\sin(\hat{\theta}_{\mathrm{r},\ell_k})\sin(\hat{\phi}_{\mathrm{r},\ell_k})\\\cos(\hat{\theta}_{\mathrm{r},\ell_k})\end{bmatrix}.\label{eq:subeq-b}
\end{align}
\end{subequations}
For each path $\ell_k$, we create a three-dimensional polyline among $\mathbf{p}_{\mathrm{t}}$, $\mathbf{p}_{\mathrm{s},\ell_k}$ and $\mathbf{p}_{\mathrm{r},k}$, as given by
\begin{equation}
\begin{aligned}
\mathbf{p}_{\mathrm{r},k}=\mathbf{p}_\mathrm{t}+c{\tau}_{\ell_k}\xi_{\ell_k}\vec{f}_{\mathrm{t,\ell_k}}+c{\tau}_{\ell_k}(1-\xi_{\ell_k})\vec{f}_{\mathrm{r},\ell_k},
\label{polyline}
\end{aligned}
\end{equation} 
where the unknown parameter $\xi_{\ell_k}\in[0,1]$ is the proportional factor of the $\ell_k$-th path along the polyline in the 3D space. It divides the 3D polyline into two segments: one from the BS to the SP and the other from the SP to the UE. For LoS paths, $\xi_{\ell_k}$ takes either 0 or 1. \eqref{polyline} can be further organized as
\begin{equation}
\mathbf{p}_{\mathrm{r},k}=\bm{u}_{\ell_k}+\xi_{\ell_k}\bm{v}_{\ell_k},
\end{equation}
where $\bm{u}_{\ell_k}=\mathbf{p}_{\mathrm{t}}-c\tau_{\ell_k}\vec{f}_{\mathrm{r},\ell_k}$ and $\bm{v}_{\ell_k}={c\tau}_{\ell_k}(\vec{f}_{\mathrm{t},\ell_k}+\vec{f}_{\mathrm{r},\ell_k})$.
A key challenge arises in accurately estimating the user's position based on the established 3D spatial polyline, while accounting for the differences in reliability among each path. To address this issue, we introduce a cost function that quantifies the cumulative error between the estimated user position and the spatial polyline. The cost function is defined as
\begin{equation} 
C(\mathbf{p}_{\mathrm{r},k}) = \sum_{\ell_k=1}^{L_k} \rho_{\ell_k}\left\| \mathbf{p}_{\mathrm{r},k} - \left( \bm{\overline{v}}_{\ell_k}+ u_{\ell_k}^\mathsf{T} (\mathbf{p}_{\mathrm{r},k}  - \bm{\overline{v}}_{\ell_k}) u_{\ell_k}\right) \right\|^2.
\label{eq:cost}
\end{equation}
where $\rho_{\ell_k}=1/({\tau_{\ell_k} c})^2$ is the weight associated with the $\ell_k$-th path and $\bm{\overline{v}}_{\ell_k}=\bm{v}_{\ell_k}/\|\bm{v}_{\ell_k}\|$. Generally, the reliability of a path is closely tied to the measurement distance. As the propagation distance increases, path loss induces elevated estimation errors and deteriorates the reliability of the path. Conversely, shorter transmission paths generally exhibit lower signal attenuation, which improves the reliability of the path~\cite{2004MitsubishiER}. This cost function essentially measures the weighted squared distance between the user’s position and the geometric constraints of each path. The inclusion of the weight $\rho_{\ell_k}$ allows the model to account for the reliability of each path, assigning greater importance to paths that are more trustworthy (with higher SNR and smaller path spread). By minimizing the cost function \eqref{eq:cost}, the least-squares solution for the UE’s position can be derived as
\begin{equation}
\hat{\mathbf{p}}_{\mathrm{r},\ell_k}=\left(\sum_{\ell_k=1}^{L_k}\rho_{\ell_k}(\mathbf{I}-\bm{\overline{v}}_{\ell_k}\bm{\overline{v}}_{\ell_k}^{\mathsf{T}})\right)^{-1}\sum_{\ell_k=1}^{L_k}\rho_{\ell_k}(\mathbf{I}-\bm{\overline{v}}_{\ell_k}\bm{\overline{v}}_{\ell_k}^{\mathsf{T}})\bm{u}_{\ell_k}.
\end{equation}
In each ping-pong round, the UE’s current position estimate $\hat{\mathbf{p}}_{\mathrm{r},k}$ can be obtained. By linking all $\hat{\mathbf{p}}_{\mathrm{r},k}$  in sequential order across ping-pong rounds, the UE’s movement trajectory can be effectively reconstructed, thus supporting accurate position and tracking in dynamic environments.

It should be noted that achieving perfect synchronization in high-precision wireless positioning systems is a highly challenging task. This problem can be estimated and compensated for using the directional consistency of signal propagation paths. Specifically, in a LoS path, the transmit and receive directions are theoretically aligned. For NLoS paths, the transmit and receive directions exhibit consistency in their projections along the LoS direction. These two key features can be used to infer and compensate for clock bias~\cite{10547696}.

\section{Computational Complexity and Convergence Behavior} \label{sec:ana}

\subsection{Time-Complexity}
The time-complexity of each major module is presented in Table~\ref{tab:time_complexity}, where $b$ is the number of channel parameters, $t$ is the number of iterations, $N_x$ refers to the grid number of any estimated parameter.  As shown in Table~\ref{tab:time_complexity}, the primary complexity lies in the construction and inversion of the FIM and SDP-based beamforming optimization. Specifically, the SDP optimization runs within an AO framework and iterates for $t$ rounds. Therefore, the overall time-complexity of the proposed algorithm is given by $\mathcal{O}\left((b N_\mathrm{c})^3+\left(t \cdot K^3N_\mathrm{c}^3\right)\right)$.

From a time-complexity perspective, the proposed method offers a key advantage: the method exhibits strong scalability in multi-antenna systems. As shown in~\eqref{eq:receivedsignals} and~\eqref{eq:C}, the dimensions of the received signal and the covariance of the received signal depend mainly on the number of OFDM symbols rather than the antenna scale, allowing the AO framework to remain efficient and hardware-friendly even for large-scale antenna arrays.

\begin{table}[t]
\renewcommand{\arraystretch}{1.8}
\caption{Time-Complexity of the Proposed Method}
\label{tab:time_complexity}
\centering
\scriptsize
\begin{tabular}{p{2.4cm}|p{2cm}|p{2.4cm}}
\hline
\textbf{Module} & \textbf{Key Parameters} & \textbf{Time-complexity} \\
\hline
Signal reconstruction and preprocessing & $K$, $N_\mathrm{c}$, $N_\mathrm{r}$, $N_\mathrm{s}$ & $\mathcal{O}(K N_\mathrm{c} N_\mathrm{r} N_\mathrm{s})$ \\
Channel estimation & $N_x$, $K$, $N_\mathrm{c}$, $N_\mathrm{r}$ & $\mathcal{O}(N^2_x+K N_\mathrm{c} N_\mathrm{r}^3)$ \\
FIM construction and inversion & $K$, $L_k$, $N_\mathrm{c}$, $N_\mathrm{r}$ & $\mathcal{O}(K L_k (N_\mathrm{c} N_\mathrm{r})^2) \newline + \mathcal{O}((b N_\mathrm{c})^3)$ \\
SDP optimization & $K$, $N_\mathrm{c}$ & $\mathcal{O}(K^3 N_\mathrm{c}^3)$ \\
AO iterations & $t$  & $\mathcal{O}(t \cdot \left(K^3 N_\mathrm{c}^3\right))$ \\
Position estimation & $K$ & $\mathcal{O}(K)$ \\
\hline
\end{tabular}
\end{table}

\subsection{Space-Complexity}
Space-complexity measures the amount of memory consumed by the proposed method, which primarily comes from storing the channel matrix and the SDP optimization matrix. Specifically, the space required for storing the channel matrix is proportional to the number of receive antennas $N_\mathrm{r}$, transmit antennas $N_\mathrm{t}$, and the number of users $K$, which results in a storage space-complexity of $\mathcal{O}(K N_\mathrm{r} N_\mathrm{t})$. On the other hand, the storage for the SDP optimization matrix involves $\mathcal{O}(K^2 N_\mathrm{c}^2)$, which grows quadratically with the number of users $K$ and the number of subcarriers $N_\mathrm{c}$.
Therefore, the overall space-complexity is dominated by these two components, and can be expressed as $\mathcal{O}(KN_\mathrm{r}N_\mathrm{t}+K^2N_\mathrm{c}^2)$.

\subsection{Convergence Behavior}
This subsection proves the convergence of the proposed AO algorithm by examining the monotonicity and boundedness of the objective function. The optimization problems \textbf{P4} and \textbf{P5} update the hybrid precoder and combiner via an SDP solver, ensuring that the objective function is either non-increasing in each iteration. Specifically, when fixing $\mathbf{W}_{k}[n]$ to optimize $\mathbf{F}[n]$, the objective function is non-increasing, and vice versa, ensuring monotonicity. Furthermore, the objective function $\frac{1}{K}\sum_{k=1}^{K} \mathrm{tr}\big(\mathbf{J}^{-1}\big(\mathbf{p}_{\mathrm{r},k} \big| \mathcal{H}_{k}\big)\big)$ is constrained by system parameters and SNR, preventing an indefinite decrease. By the monotone convergence theorem, a monotonic and bounded sequence is guaranteed to converge.

\section{Simulation Analysis} \label{sec:simu}
\subsection{Parameter Configuration}
This section presents a series of results to evaluate the performance of the proposed ping-pong positioning scheme. Root mean square error (RMSE) is chosen for the positioning accuracy metric.

We set $f_\mathrm{c} = 28~\mathrm{GHz}$, $\lambda_\mathrm{c} = 10 e^{-3}~\mathrm{m}$ and $\Delta f = 15~\mathrm{kHz}$. The noise spectral density is set to -114 dBm/MHz.

The benchmark schemes are presented for comparison:
\begin{enumerate}
\item VB: A virtual BS (VB) method proposed in~\cite{10063228} for addressing the NLoS positioning problem.
\item Relay: An active relay-assisted positioning method, as discussed in~\cite{9834216}.
\item ToF: A classic time of flight (ToF)-based positioning method is presented in~\cite{9152004}.
\item Rank-1 EFIM: A method integrates the contribution of each NLoS component into a separate rank-one matrix within the extended FIM (EFIM) and utilizes these components for positioning, which is proposed in~\cite{8515231}.
\item R-angle: A positioning method from~\cite{10741287} that employs rank estimation to analyze scattering paths.
\item PHD: A map-based multi-model probability hypothesis density (PHD) filter method for positioning in~\cite{9032328}.
\end{enumerate}

\subsection{Simulation Analysis}

\begin{figure*}[ht!]
    \captionsetup{font=footnotesize}
    \centering
    \begin{subfigure}[b]{0.3\textwidth}
        \centering
        \includegraphics[width=\linewidth, height=5.5cm]{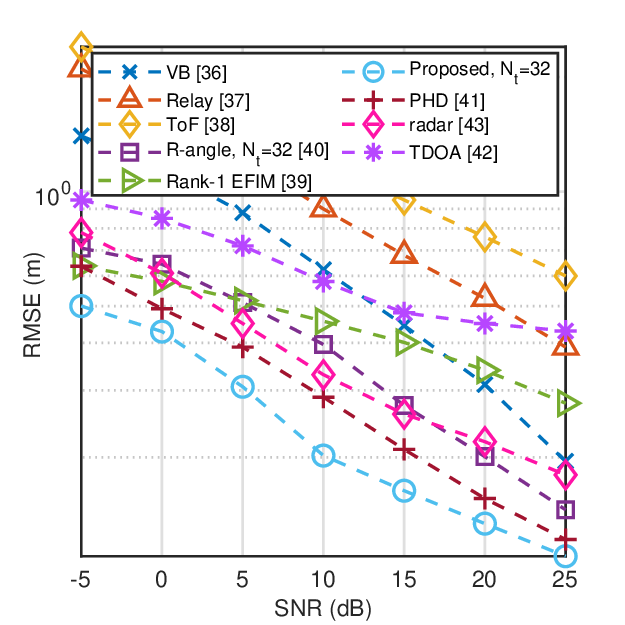}
        \caption{}
    \end{subfigure}
    \hfill
    \begin{subfigure}[b]{0.3\textwidth} 
        \centering
        \includegraphics[width=\linewidth, height=5.5cm]{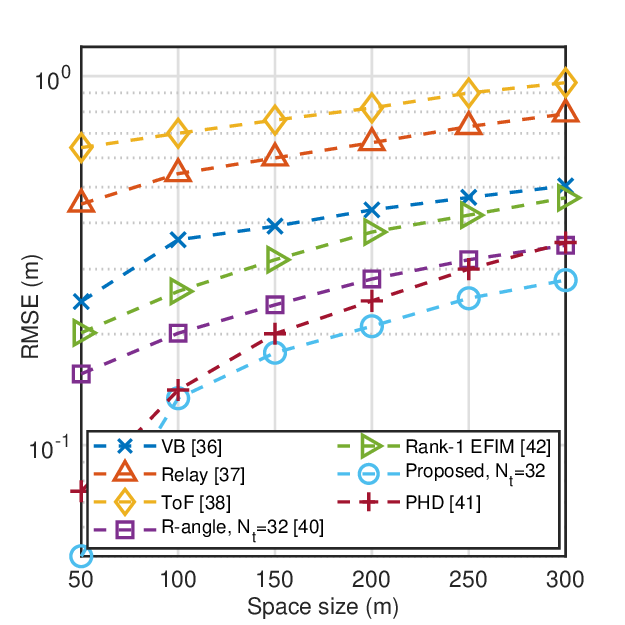}
        \caption{}
    \end{subfigure}
    \hfill
    \begin{subfigure}[b]{0.3\textwidth}
        \centering
        \includegraphics[width=\linewidth, height=5.5cm]{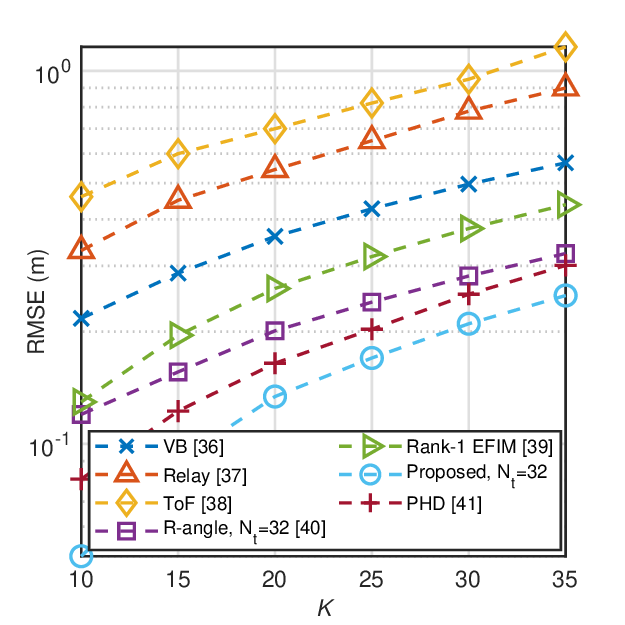}
        \caption{}
    \end{subfigure}
    \caption{Comparison for different positioning methods: (a) RMSE vs. SNR, where $K=20$, Space size\:=\:100\;m; (b) RMSE vs. Space size, where $K=20$, SNR\:=\:20\;dB; (c) RMSE vs. $K$, where Space size\:=\:100\;m, SNR\:=\:20\;dB.}
    \label{simu:methods}
\end{figure*}

In Fig.~\ref{simu:methods}, we compare the performance of various positioning algorithms.
Fig.~\ref{simu:methods}(a) illustrates the achievable positioning RMSE as a function of SNR. It is seen that the RMSE of all methods decreases significantly as the SNR increases from -5 dB to 25 dB. 
The proposed method ($N_\mathrm{t}=32$) demonstrates superior performance in low-SNR regions, with its RMSE significantly lower than classical methods such as VB, RELAY, and ToF. 
The proposed method consistently achieves lower RMSE than both TDoA and radar-based positioning across various SNR levels. 

As shown in Fig.~\ref{simu:methods}(b), the RMSE of all methods increases with the spatial dimension, ranging from 50 m to 300 m. This trend arises from a decline in geometric precision as the spatial scale grows, making the position more challenging. In smaller spatial areas, the proposed method and the PHD method achieve centimeter-level accuracy, whereas other methods remain at the decimeter level. RELAY and ToF experience a sharp increase in RMSE with larger spatial dimensions, reaching nearly 1 m at 300 m, while the proposed method exhibits a smaller RMSE increase, demonstrating superior scalability in extended spatial domains.

As depicted in Fig.~\ref{simu:methods}(c), the RMSE of all methods grows with increasing $K$. This increase is primarily due to additional multi-user interference. For smaller $K$, the performance of R-angle ($N_\mathrm{t}=32$) methods is comparable to the proposed method, but the other methods exhibit greater fluctuations as the user number rises. The proposed method maintains robust performance in high user-density scenarios and effectively mitigates multi-user interference.

\begin{figure}[ht!] 
    \captionsetup{font=footnotesize}
    \centering
    \begin{subfigure}[b]{0.48\textwidth} 
        \centering
        \includegraphics[width=3.5in]{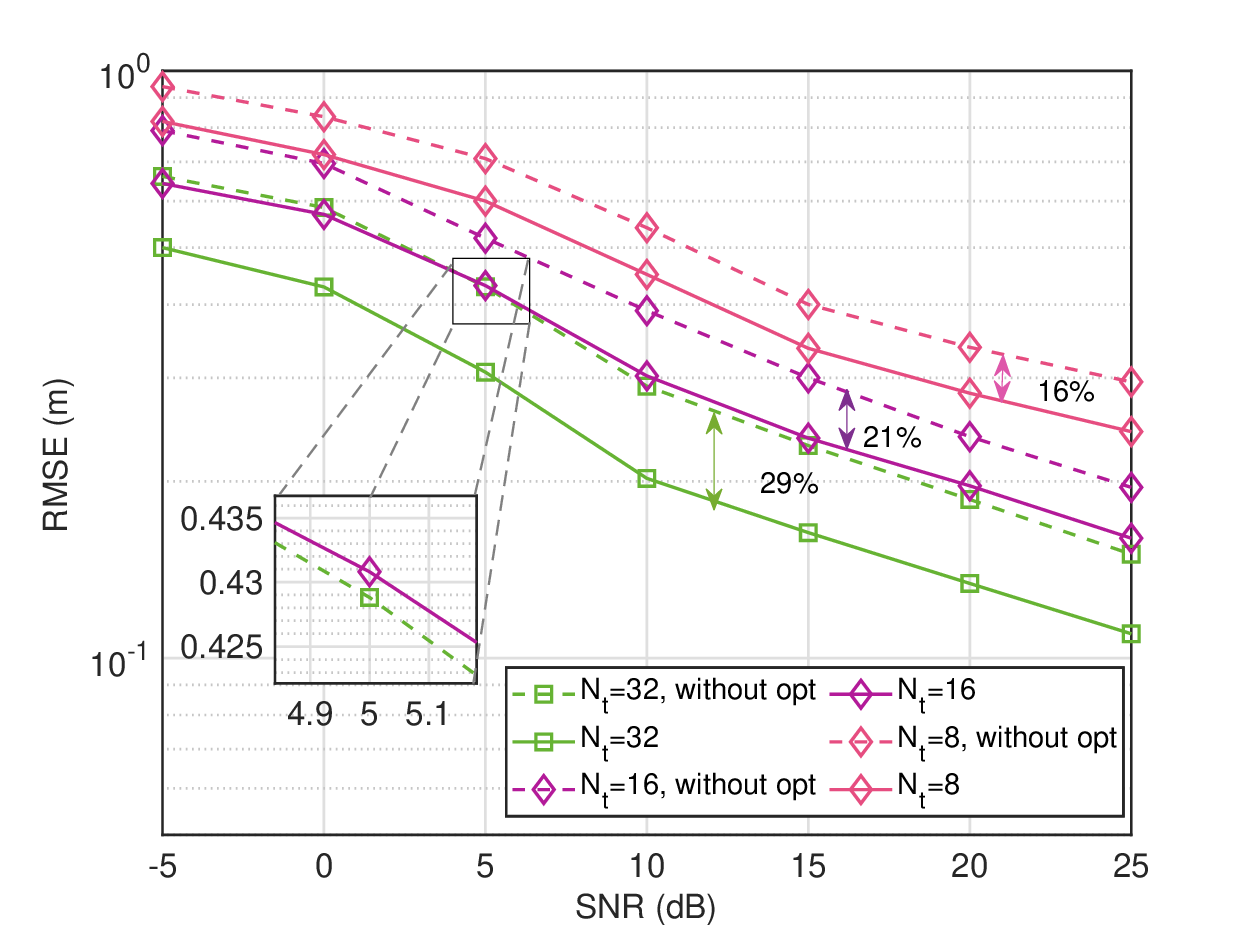} 
        \caption{}
    \end{subfigure}
    \hfill
    \begin{subfigure}[b]{0.48\textwidth} 
        \centering
        \includegraphics[width=3.5in]{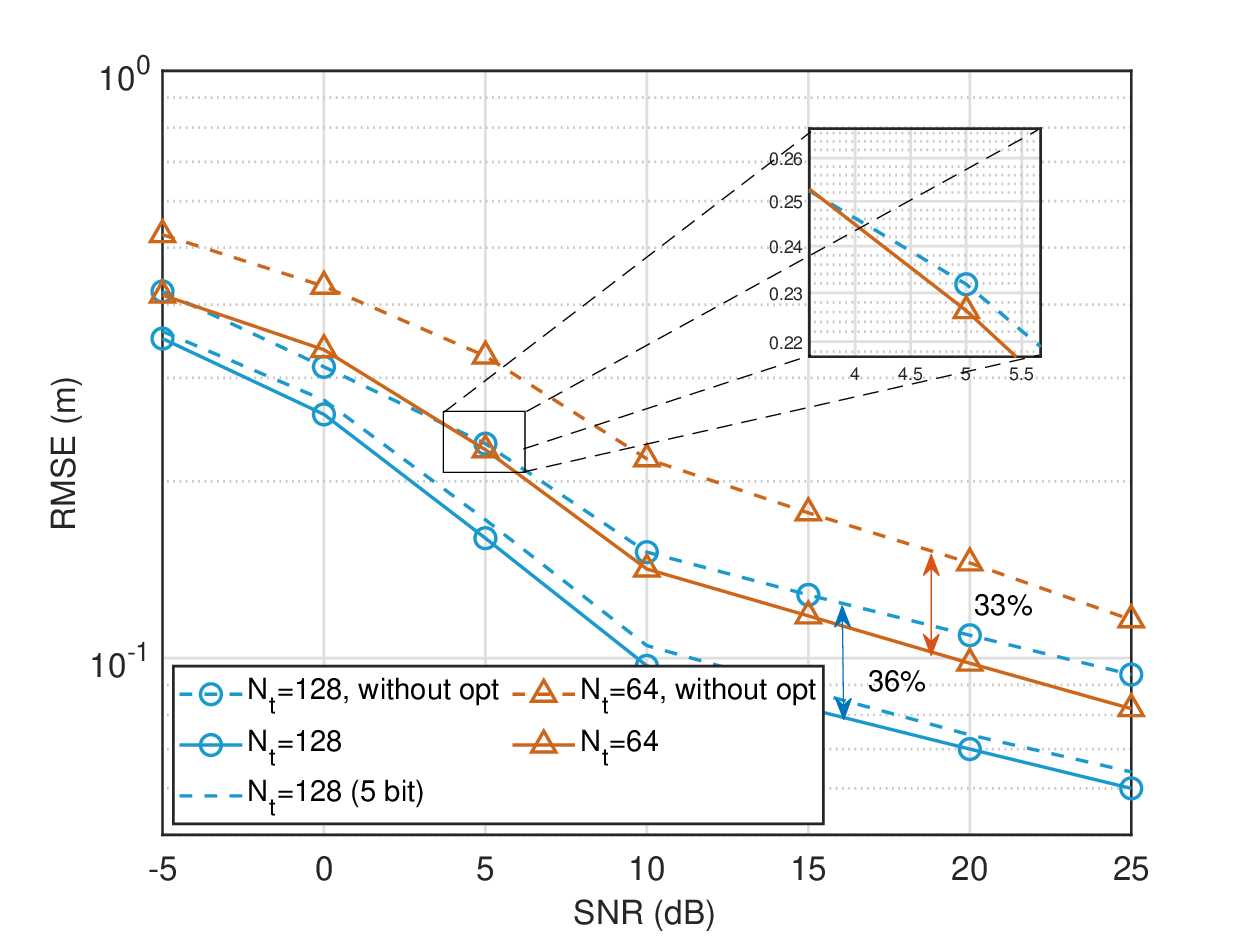} 
        \caption{}
    \end{subfigure}
    \caption{The impact of optimization for MSCPEB on RMSE under different antenna configurations, where $K=20$ and space size\:=\:100\;m. The performance gaps are annotated as percentages.}
    \label{simu:opt vs rmse}
\end{figure}

Fig.~\ref{simu:opt vs rmse} investigates the impact of optimizing the MSCPEB on positioning performance, comparing positioning schemes with and without MSCPEB optimization. We see that optimization leads to significant improvements in positioning accuracy, especially at higher SNR levels. On the other hand, the comparison between large antenna arrays and smaller ones is shown in Fig.~\ref{simu:opt vs rmse}. In Fig.~\ref{simu:opt vs rmse}(a), larger arrays ($N_\mathrm{t}=64$ and $N_\mathrm{t}=128$) show substantial performance gains from optimization, especially at medium to high SNR levels. For $N_\mathrm{t}=128$, the RMSE is reduced by up to 36\%, achieving centimeter-level accuracy at 25dB. Similarly, a 33\% improvement is observed for $N_\mathrm{t}=64$. Fig.~\ref{simu:opt vs rmse}(b) presents results for smaller arrays ($N_\mathrm{t}=32, 16, 8$), where optimization still offers moderate benefits, such as a 29\% reduction for $N_\mathrm{t}=32$. However, the overall RMSE remains relatively high, especially for $N_\mathrm{t}=8$, where the improvement is limited to around 16\%. These results indicate that the proposed method is more effective in large-scale antenna arrays. Furthermore, accounting for the finite resolution of practical digital phase shifters, we evaluate the positioning RMSE using 5-bit-resolution digital phase shifters under an SNR of 15 dB with 128 antennas. As demonstrated, the 5-bit configuration achieves an RMSE closely approaching that of the ideal infinite-resolution case, thus achieving centimeter-level accuracy. This performance preservation stems from the hybrid beamforming architecture in the proposed framework. The analog phase shifters provide theoretically infinite resolution. This property effectively compensates for the precision limitations of digital phase shifters, simultaneously demonstrating the practical value of adopting hybrid beamforming in the proposed framework.

\begin{figure}[h]
  \captionsetup{font=footnotesize}
  \centering
  \includegraphics[width=3.5in]{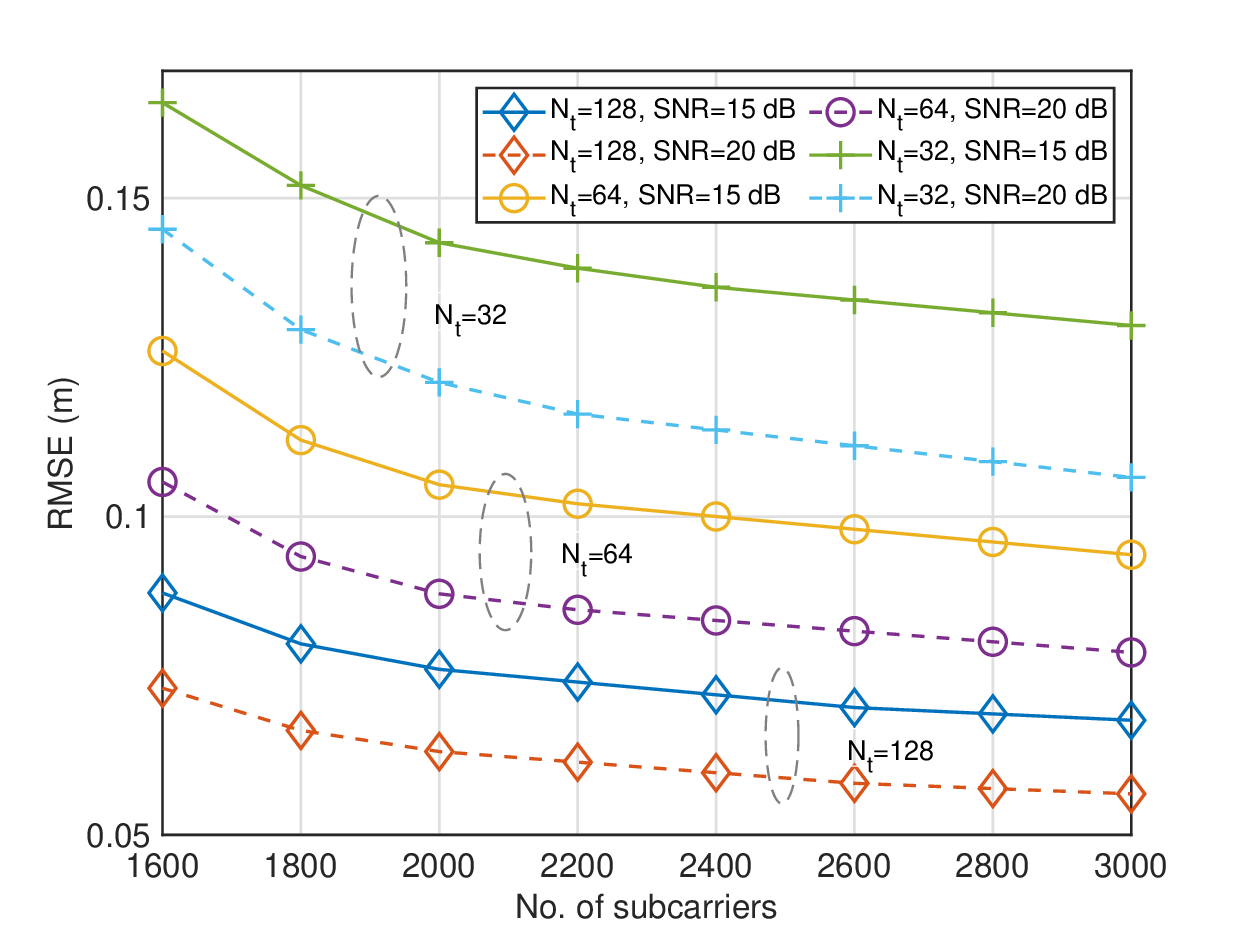}
  \caption{Positioning RMSE vs. number of subcarriers, with $K=20$ and space size\:=\:100\;m.}
  \label{simu:subcarriers vs rmse}
\end{figure}

Fig.~\ref{simu:subcarriers vs rmse} explores the effect of the number of subcarriers on RMSE, illustrating that an increase in system DoF enhances positioning accuracy. As the number of subcarriers increases, the improved RMSE validates that the additional DoF contributes to improved estimation accuracy. This demonstrates that the proposed method effectively exploits the collaborative gain across multiple subcarriers, leveraging the additional degrees of freedom to improve the positioning accuracy in wideband systems. When the number of subcarriers exceeds 2400, the RMSE decreases minimally and stabilizes, indicating that the system performance is approaching its maximum potential.

\begin{figure}[t]
  \captionsetup{font=footnotesize}
  \centering
  \includegraphics[width=3.5in]{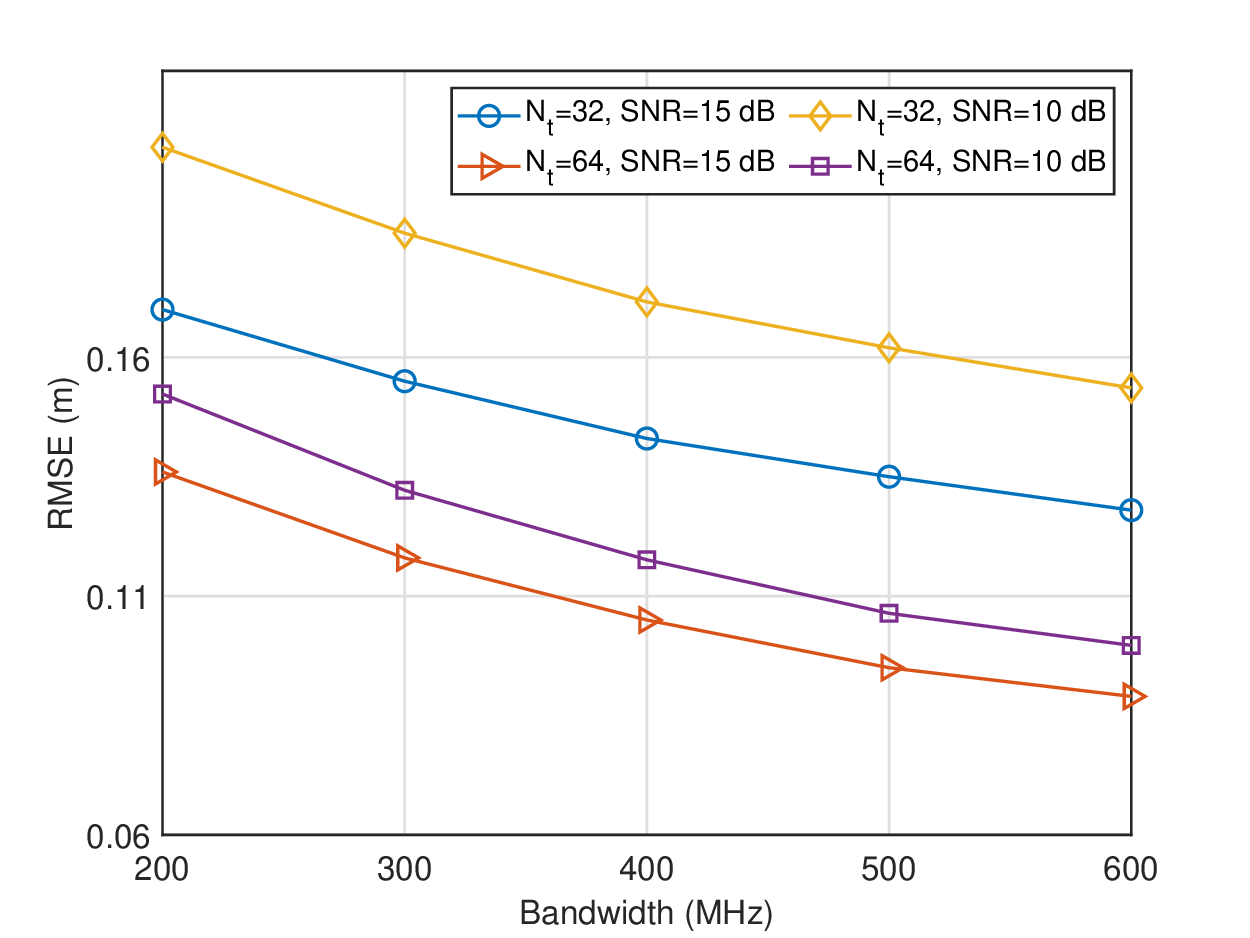}
  \caption{Positioning RMSE vs. bandwidth, with SNR\:=\:10\:dB and 15\:dB.}
  \label{simu:bandwidth}
\end{figure}

As shown in Fig.~\ref{simu:bandwidth}, the RMSE of positioning decreases slightly with increasing bandwidth. According to $\Delta f =W_\mathrm{c}/N_\mathrm{c}$, with $N_\mathrm{c}$ is fixed, increasing the system bandwidth results in larger subcarrier spacing. As derived in~\eqref{eq:correlation1}, the subcarrier correlation factor grows exponentially with bandwidth.  A larger correlation factor implies lower inter-subcarrier correlation, meaning that the subcarrier can provide greater positioning-related observation information (i.e., FIM). Therefore, both MSCPEB and the positioning error are reduced. Conversely, reducing the bandwidth narrows the differences in the FIM across subcarriers, leading to a more balanced contribution from all subcarriers to the positioning outcome.

\begin{figure}[t]
  \captionsetup{font=footnotesize}
  \centering
  \includegraphics[width=3.5in]{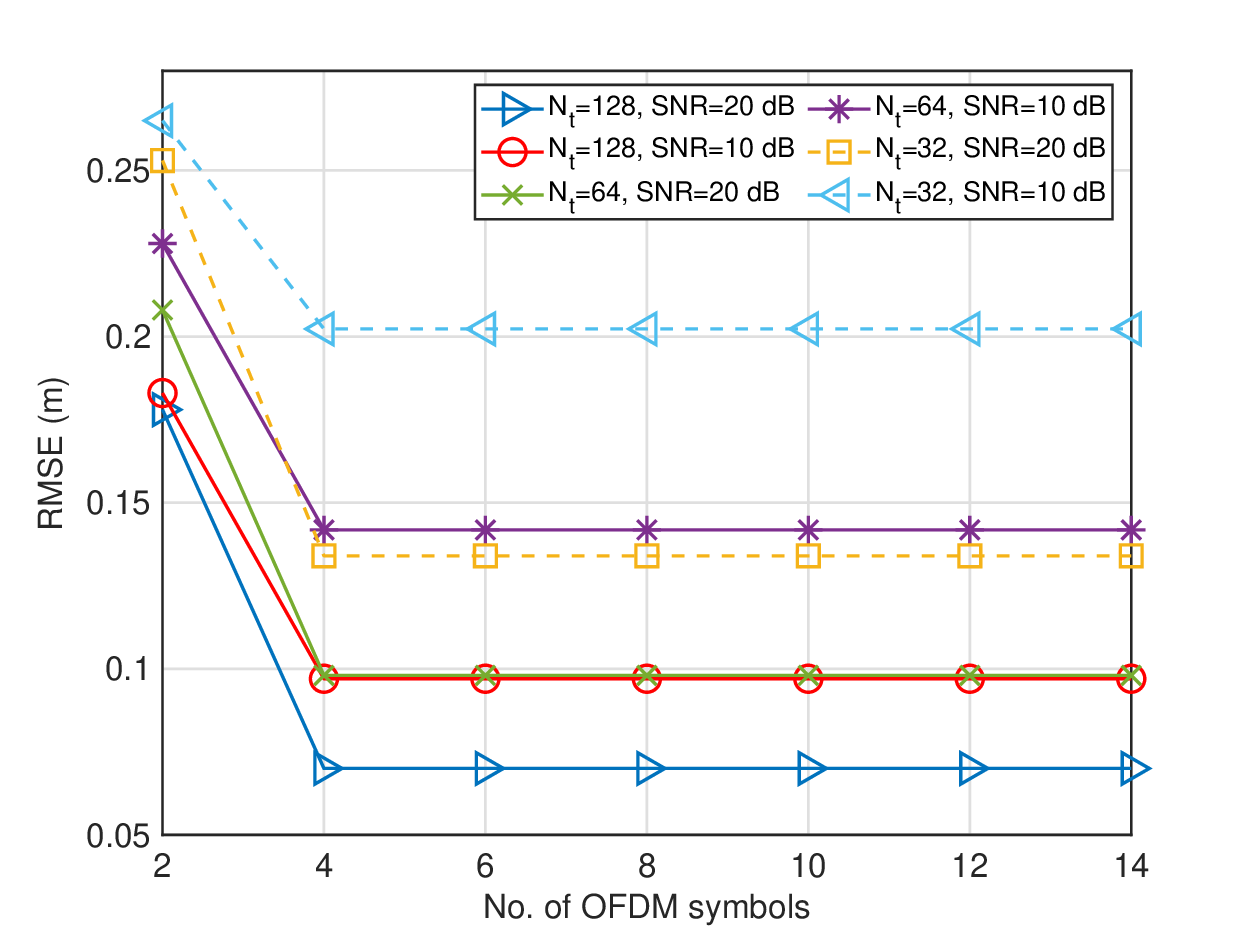}
  \caption{Positioning RMSE vs. number of OFDM symbols, with $K=20$ and space size\:=\:100\;m.}
  \label{simu:ofdm vs rmse}
\end{figure}

Fig.~\ref{simu:ofdm vs rmse} plots the impact of transmitted OFDM symbols on positioning accuracy.  It is seen that positioning accuracy stabilizes once the number of OFDM symbols exceeds 4. The proposed method can achieve high-accuracy positioning with only a small number of OFDM symbols. This highlights the efficiency and robustness of the algorithm, as it maintains superior performance even with a limited number of symbols. In 5G NR systems, each slot typically contains 14 symbols. The proposed method approaches its performance limit using only approximately one-quarter of the slot resources. Consequently, this efficient use of spectrum resources ensures robust positioning with minimal overhead, making our method highly suitable for resource-constrained wideband systems.

\begin{figure}[t]
  \captionsetup{font=footnotesize}
  \centering
  \includegraphics[width=3.5in]{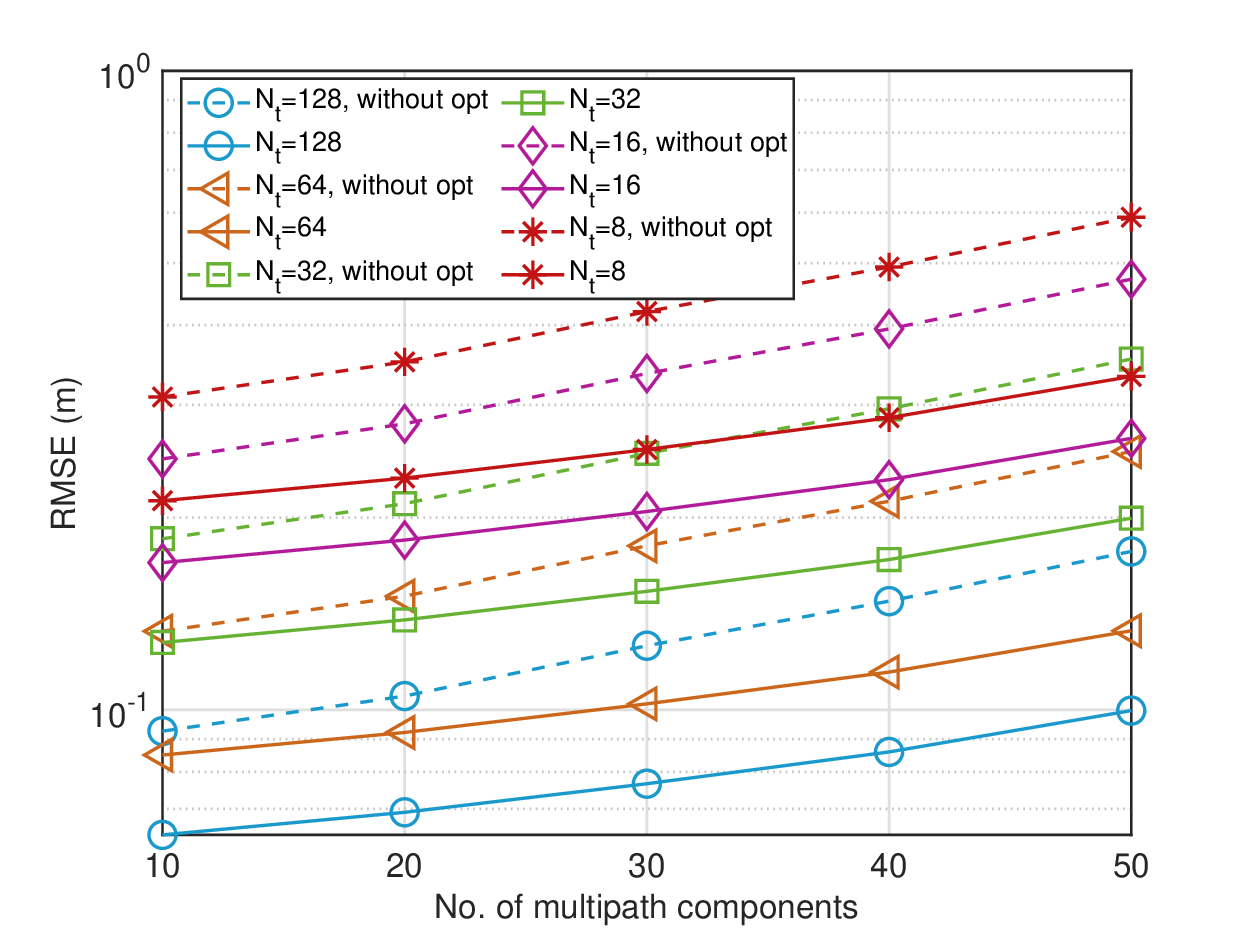}
  \caption{Positioning RMSE vs. number of total multipath components, with $K=10$ and space size\:=\:100 m.}
  \label{simu:multipath vs rmse}
\end{figure}

Fig.~\ref{simu:multipath vs rmse} illustrates the impact of varying levels of multipath interference on RMSE across different antenna configurations. The MSCPEB-optimized schemes consistently outperform their non-optimized counterparts, with the performance gap widening significantly as the number of multipath components increases (e.g., at 40 or 50 components). MSCPEB-optimized schemes effectively harness multi-dimensional information, markedly improving robustness in challenging conditions.

\begin{figure}[t]
  \captionsetup{font=footnotesize}
  \centering
  \includegraphics[width=3.5in]{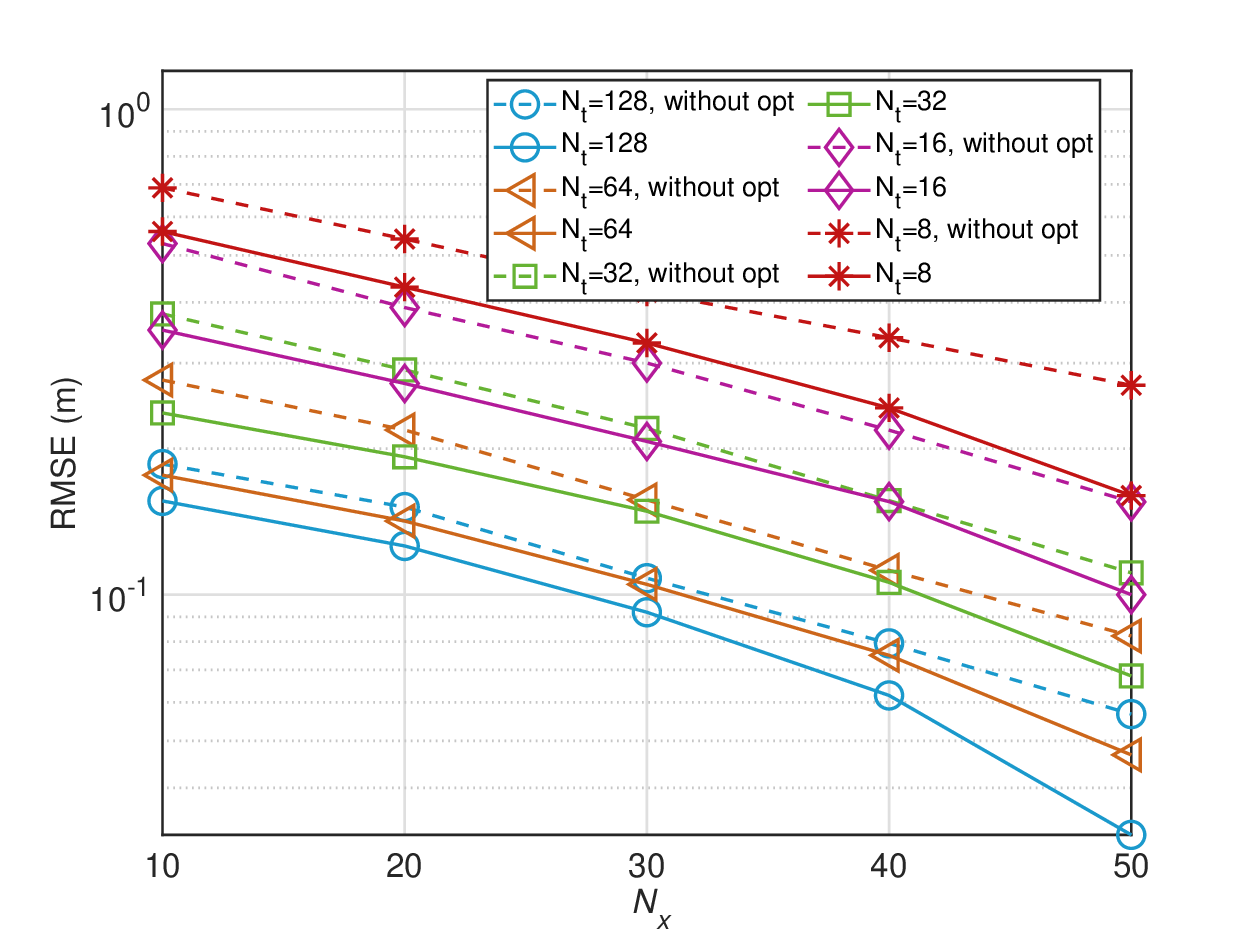}
  \caption{Positioning RMSE vs. grid number of uninformed search, where $K=20$ and space size\:=\:100\;m. Assuming all parameters have the same grid number, $N_x$ refers to the grid number of any estimated parameter.}
 \label{simu:nx vs rmse} 
\end{figure}

Fig.~\ref{simu:nx vs rmse} demonstrates the impact of uninformed estimation accuracy on the robustness of the positioning method. Increasing the number of uninformed search grids $N_x$ leads to a significant reduction in RMSE for all schemes, highlighting the critical role of uninformed estimation precision in overall positioning performance. When $N_x$ is small, the initial uninformed estimation deviates more significantly from the true parameters, adversely affecting final positioning accuracy. However, the MSCPEB-optimized scheme (solid lines) consistently outperforms the non-optimized scheme (dashed lines), with the performance gap widening as $N_x$ increases. This demonstrates that, although uninformed estimation deviations affect positioning accuracy, the MSCPEB optimization method helps mitigate this impact, enhancing the positioning robustness.

\begin{figure}[t]
  \captionsetup{font=footnotesize}
  \centering
  \includegraphics[width=3.5in]{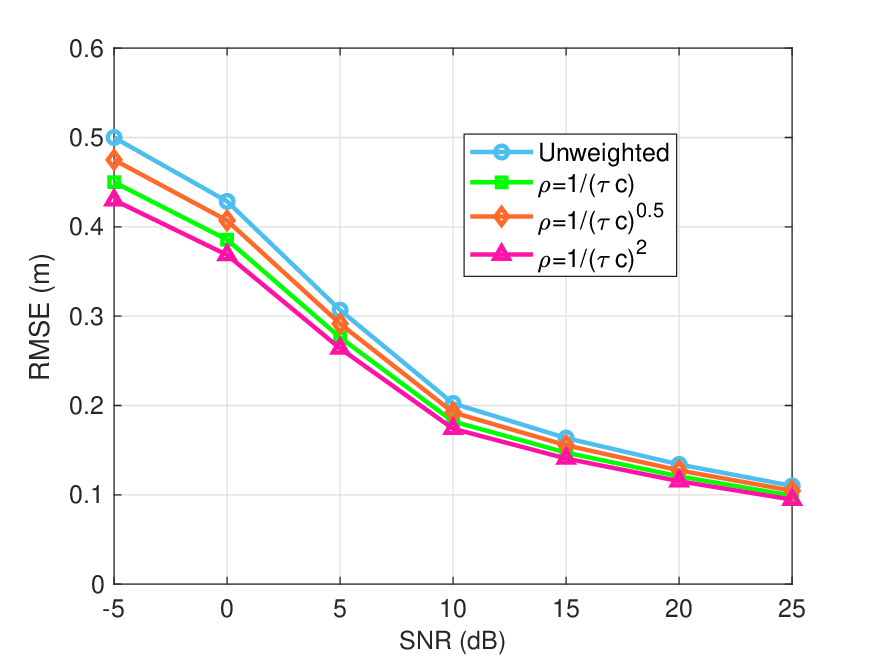}
  \caption{Positioning RMSE vs. SNR for different path weighting schemes.}
  \label{simu:weight}
\end{figure}

Fig.~\ref{simu:weight} illustrates the relationship between SNR and RMSE for various path weighting schemes. Specifically, we compare the performance of unweighted estimation with three different weighting functions, which are based on the path length.  All weighting schemes achieve lower RMSE compared to the unweighted case, demonstrating that path weights can improve the positioning accuracy. Notably, the weighting function that is inversely proportional to the square of the path length consistently delivers the best performance across the entire SNR.

\begin{figure}[t]
  \captionsetup{font=footnotesize}
  \centering
  \includegraphics[width=3.5in]{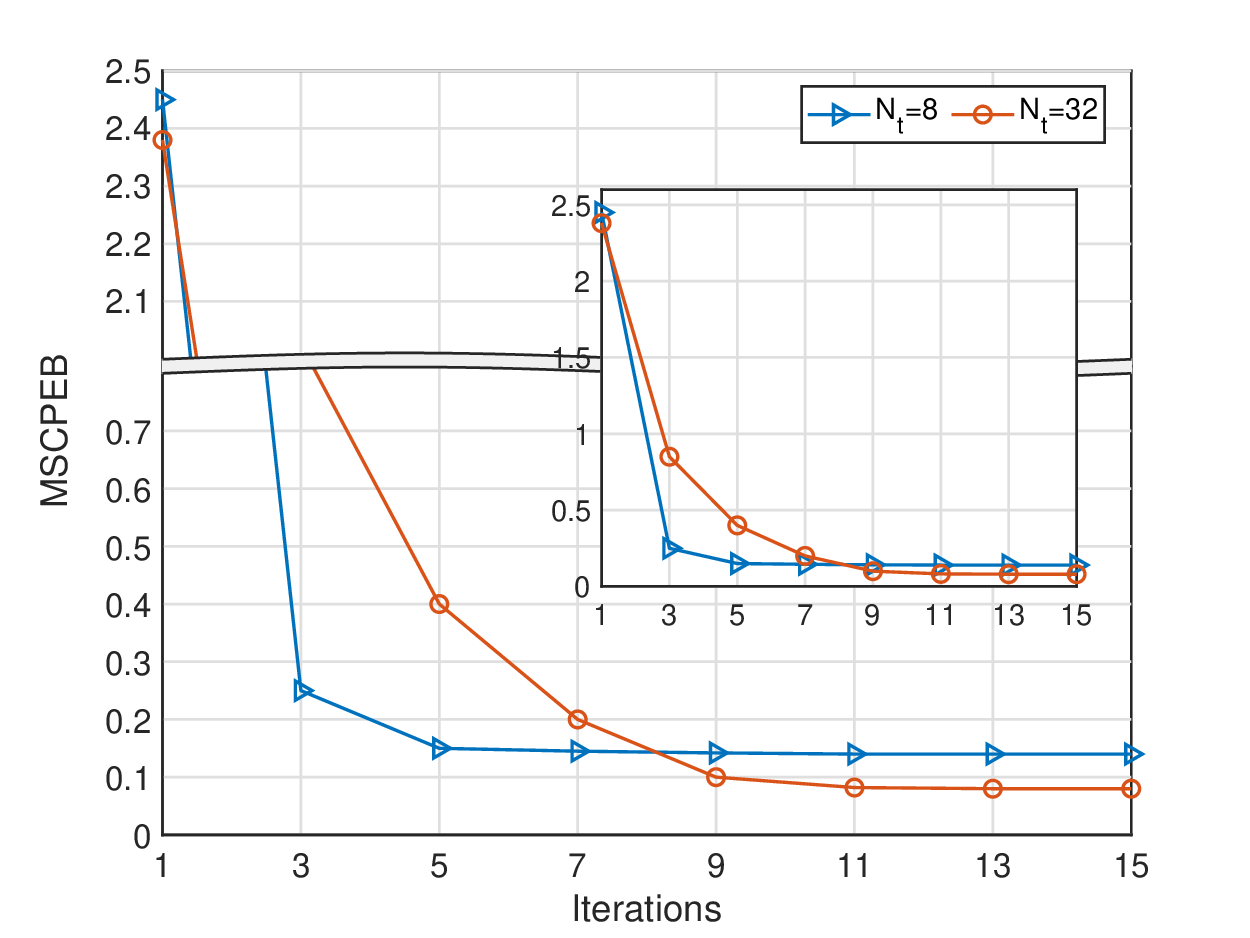}
  \caption{Convergence speed of MSCPEB, with SNR\:=\:15\:dB: detailed view (inset) and overall trend (main plot).}
  \label{simu:Converging}
\end{figure}

Fig.~\ref{simu:Converging} illustrates the convergence behavior of MSCPEB under an SNR of 15~dB. The main plot provides a detailed view of the convergence process by truncating the vertical axis. With 8 antennas, the algorithm converges rapidly within 4--6 iterations, whereas with 32 antennas, convergence takes approximately 11--13 iterations. The inset plot presents the overall convergence trend, which confirms the scalability of the proposed method in multi-antenna systems. Although the convergence speed slightly decreases as the number of antennas increases, our algorithm still converges within a reasonable number of iterations. In terms of computational efficiency, the time required for a single outer loop (AO) is approximately 90~ms (Intel i7-8700~CPU, 3.20~GHz, and 1.19~GB of available RAM). The total time for completing one full beamforming calculation ranges from 0.8~s to 1.2~s. In urban environments, typical walking speeds range from 1.0 to 1.5 m/s, while vehicle speeds range from 20 to 50 km/h (5.6 to 13.9 m/s). These displacements remain within the spatial coherence distance of mmWave systems, where a typical beamwidth (3--15\degree) can effectively cover movements up to 25--50~m~\cite{TR38901}. Therefore, the proposed algorithm meets the real-time requirements for urban environments.

\begin{figure}[t]
    \captionsetup{font=footnotesize}
    \centering
    \begin{subfigure}[b]{0.48\textwidth} 
        \centering
        \includegraphics[width=3.5in]{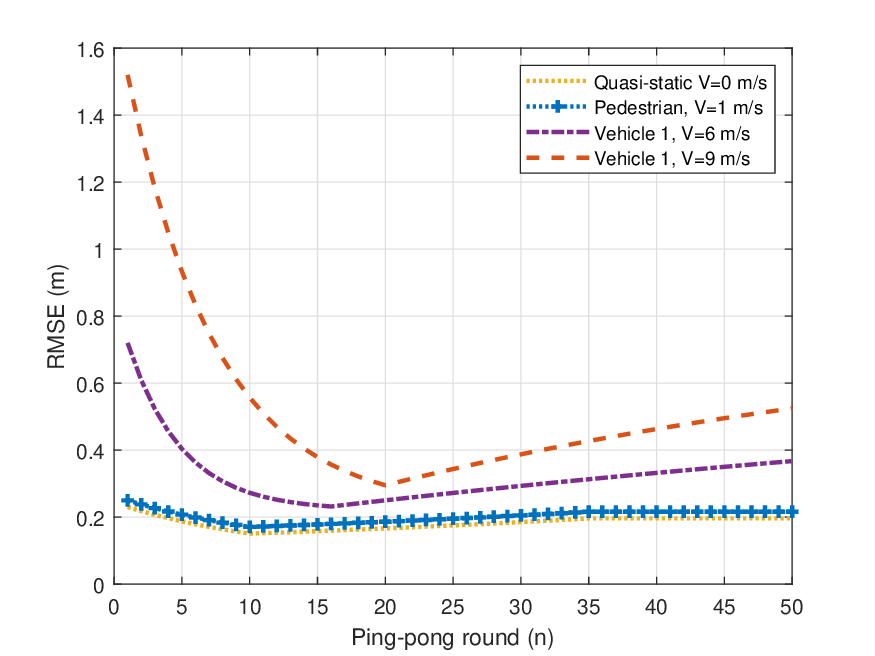} 
        \caption{}
        \label{simu:speed}
    \end{subfigure}
    \hfill
    \begin{subfigure}[b]{0.48\textwidth} 
        \centering
        \includegraphics[width=3.5in]{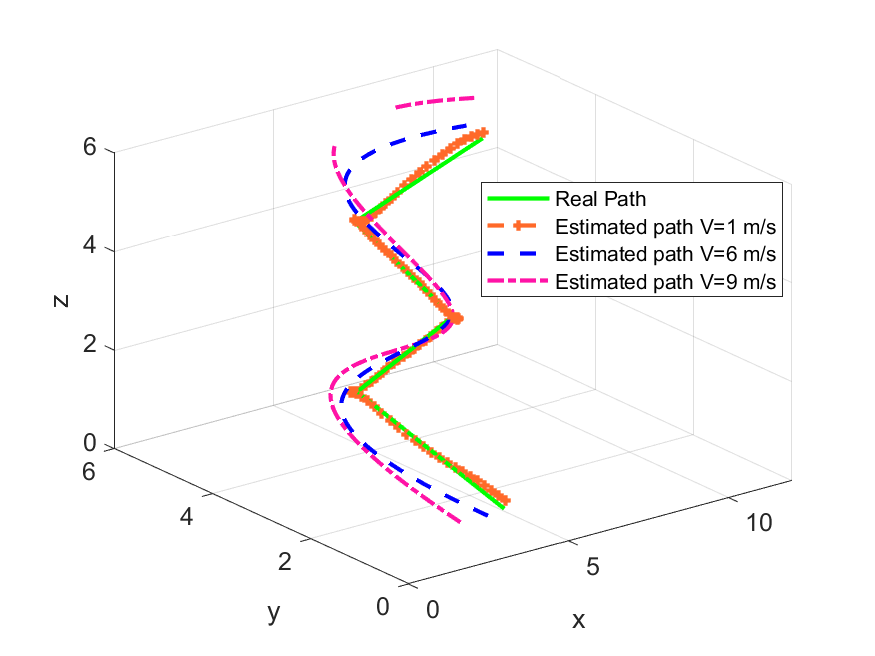} 
        \caption{}
        \label{simu:trajectory.eps}
    \end{subfigure}
    \caption{The impact of dynamic user mobility on the robustness of the proposed method.}
    \label{simu:dynamic}
\end{figure}

Fig.~\ref{simu:dynamic} presents the positioning RMSE and trajectory estimates. The proposed positioning method is tailored for slow-moving users in urban environments, assuming quasi-static channel characteristics during the ping-pong process. To evaluate its robustness under dynamic mobility, we adopt Jakes' model~\cite{Jakes1993}, which simulates the fading process induced by user movement in time-varying mobile channels.  As shown in Fig.~\ref{simu:dynamic}(a), for dynamic users with lower velocities (e.g., 1 m/s and 6 m/s), the RMSE remains relatively stable, indicating that the method is robust to moderate mobility. After several rounds of the ``ping-pong'' process, the proposed method progressively reduces positioning error, enhancing estimates from coarse to fine. However, as the number of rounds increases, error accumulation becomes more evident, particularly at higher speeds, where Doppler shifts significantly degrade position estimation. Fig.~\ref{simu:dynamic}(b) depicts the true trajectory and estimated trajectories for users at the different speeds. The trends in Fig.~\ref{simu:dynamic}(b) mirror those in Fig.~\ref{simu:dynamic}(a). While the method excels under moderate mobility, it faces increasing challenges in high-speed scenarios due to cumulative estimation errors.

\begin{figure}[h]
  \captionsetup{font=footnotesize}
  \centering
  \includegraphics[width=3.5in]{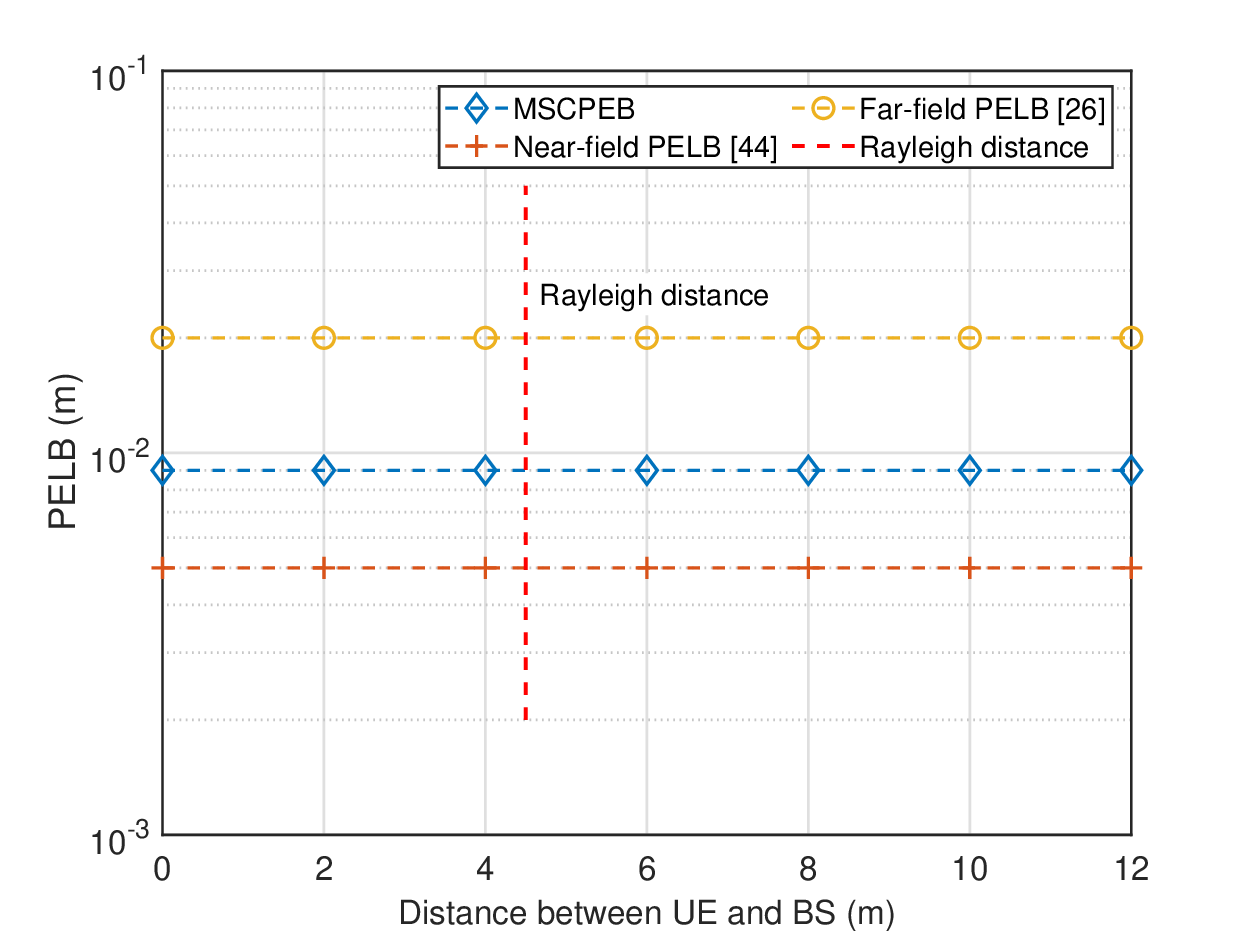}
  \caption{MSCPEB vs. single-subcarrier PELB, with SNR\:=\:15\:dB.}
\label{simu:rayleigh}
\end{figure}

Fig.~\ref{simu:rayleigh} compares the proposed MSCPEB with single-subcarrier near-field \cite{10988576} and single-subcarrier far-field PELBs \cite{10370743}. As depicted, the MSCPEB maintains invariance as the user crosses the Rayleigh distance. The underlying reason is that the MSCPEB remains rooted in far-field array responses given by \eqref{eq:arrayres1} and \eqref{eq:arrayres2}. In other words, it fails to capture the near-field-specific spatial phase curvature information. Consequently, under a fixed system model, the MSCPEB varies solely with the SNR. While consistently higher than the single-subcarrier near-field PELB, the MSCPEB is lower than the single-subcarrier far-field PELB. These findings validate the efficacy of the MSCPEB: it significantly outperforms conventional far-field PELB while achieving near-field positioning accuracy (on the order of $10^ {-3}$ m) within the near-field region. This performance stems from the MSCPEB's exploitation of the inherent rich frequency of the OFDM system, which partially compensates for the absence of near-field spatial phase curvature information in far-field channel models.

\begin{figure}[t]
  \captionsetup{font=footnotesize}
  \centering
  \includegraphics[width=3.5in]{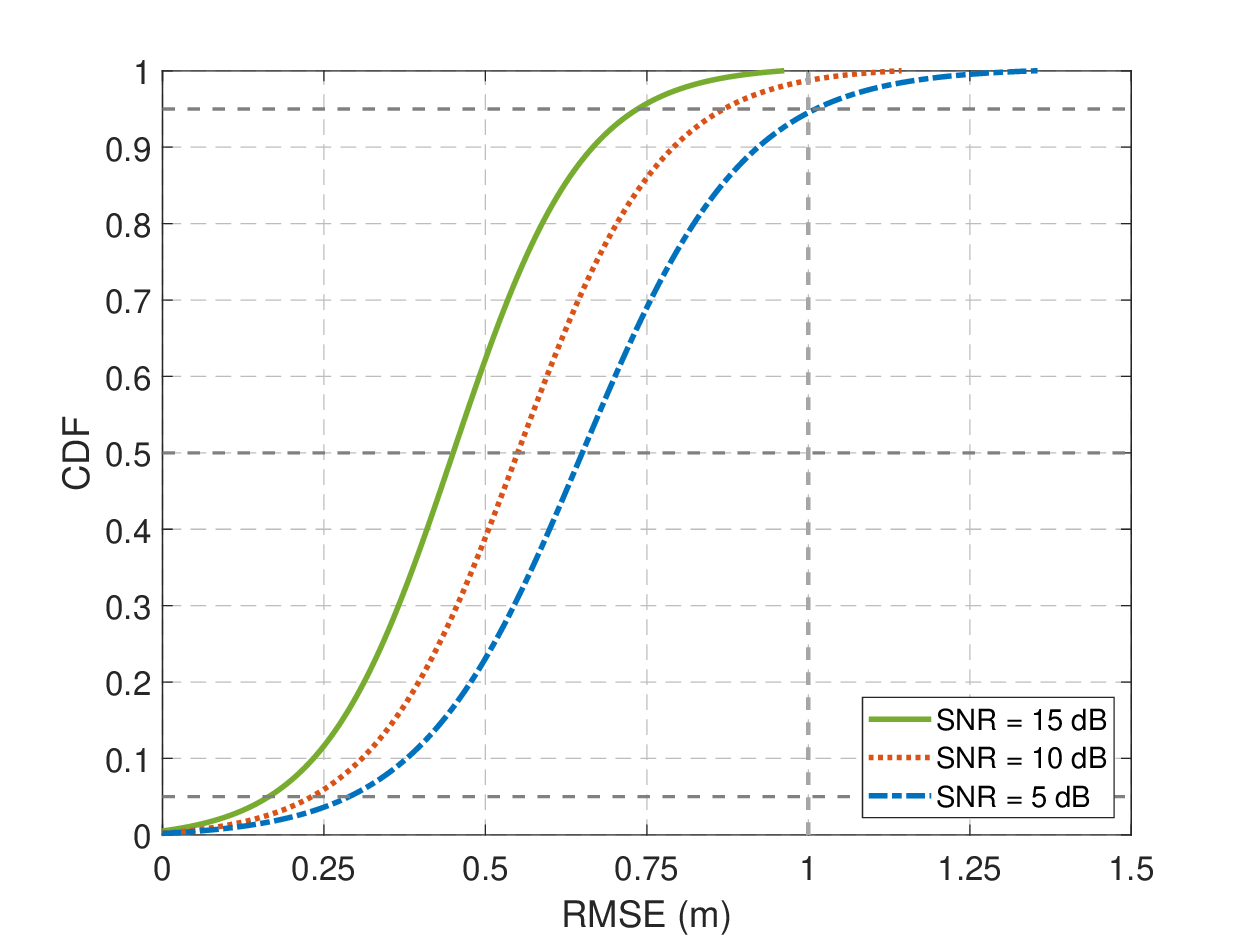}
  \caption{The cumulative distribution of positioning RMSE, with $K=20$.}
\label{simu:CDF}
\end{figure}

Fig.~\ref{simu:CDF} shows the cumulative distribution function (CDF) of per-user positioning RMSE under varying SNR levels. As SNR increases, the positioning accuracy improves significantly. Even at SNR = 5 dB, sub-meter-level accuracy remains achievable for 95\% of users. These results exceed the 3GPP requirements for positioning accuracy, achieving horizontal accuracy below 10 meters and vertical accuracy below 3 meters, confirming that the proposed positioning method does not sacrifice performance for edge users~\cite{3gpp2019tr38855}.

\section{Conclusion} \label{sec:conc}
This paper proposes a device-based PELB-driven ping-pong position framework, which enhances the accuracy of dynamic UE  positioning and tracking through beam scheduling in wideband mmWave systems. In the ping-pong framework, the UE (i.e., positioning target) can actively transmit signals and is equipped with multiple antennas. By alternately transmitting and receiving adaptive beamformed signals between the BS and UEs, the framework iteratively refines channel parameters and position estimates via an uninformed-to-informed approach, without requiring prior knowledge of CSI or UE positions. To harness the rich features from received signals, we propose a multi-dimensional information fusion method, including MSCPEB and multipath collaborative positioning, to enhance both precision and robustness in positioning performance. Simulations confirm that the optimized beam configuration improves positioning accuracy by at least 16\%. The proposed method requires only a quarter of the timeslot resources, demonstrating high resource efficiency and low latency.

\begin{appendices}
\section{Proof of Lemma \ref{lemma:PDF}.}\label{appendix3}
Given the decomposition of the $n$-th subcarrier interference signal received by the $k$-th user into two parts, $\mathbf{s}_{i,1}[n]$ and $\mathbf{s}_{i,2}[n]$, the interference affecting the $k$-th user on the $n$-th subcarrier $\mathbf{B}_{k}[n]$ is
\begin{align}
\mathbf{B}_{k}[n]=&\sum_{i\neq k}\mathbf{W}_{\mathrm{BB},k}^{\mathsf{H}}[n]\mathbf{W}_{\mathrm{RF},k}^{\mathsf{H}}\mathbf{H}_{k}[n]\mathbf{F}_{\mathrm{RF}}\mathbf{F}_{\mathrm{BB},i}[n]\nonumber\\
&\times (\mathbf{s}_{i,1}[n]+\mathbf{s}_{i,2}[n]).
\label{eq:additive1}
\end{align}
Through the distributive matrix multiplication, we have
\begin{align}
\mathbf{B}_{k}[n] = &\sum_{i\neq k}\mathbf{W}_{\mathrm{BB},k}^{\mathsf{H}}[n]
\mathbf{W}_{\mathrm{RF},k}^{\mathsf{H}}\mathbf{H}_{k}[n]
\mathbf{F}_{\mathrm{RF}}\mathbf{F}_{\mathrm{BB},i}[n]\mathbf{s}_{i,1}[n] \nonumber\\
&+ \sum_{i\neq k}\mathbf{W}_{\mathrm{BB},k}^{\mathsf{H}}[n]
\mathbf{W}_{\mathrm{RF},k}^{\mathsf{H}}\mathbf{H}_{k}[n]
\mathbf{F}_{\mathrm{RF}}\mathbf{F}_{\mathrm{BB},i}[n]\mathbf{s}_{i,2}[n].
\label{eq:additive2}
\end{align}
Thus, we have
\begin{align}
\mathbf{B}_k[n]=\mathbf{B}_{k,1}[n]+\mathbf{B}_{k,2}[n].
\label{eq:additive3}
\end{align}
\eqref{eq:additive3} illustrates the additive closure property.
Next, we verify the closure property under scalar multiplication. Considering the input signal $a\mathbf{s}_{i}[n]$, the interference is expressed by
\begin{align}
\mathbf{B}(a s_i[n])=\sum_{i\neq k}\mathbf{W}_{\mathrm{BB},k}^{\mathsf{H}}[n]\mathbf{W}_{\mathrm{RF},k}^{\mathsf{H}}\mathbf{H}_{k}[n]\mathbf{F}_{\mathrm{RF},i}\mathbf{F}_{\mathrm{BB},i}[n](a\mathbf{s}_{i}[n]).
\label{eq:additive3_1}
\end{align}
Based on the closure property of matrix multiplication under scalar multiplication, \eqref{eq:additive3_1} is rewritten as
\begin{equation}
\mathbf{B}(a s_i[n])=a\sum_{i\neq k}\mathbf{W}_{\mathrm{BB},k}^{\mathsf{H}}[n]\mathbf{W}_{\mathrm{RF},k}^{\mathsf{H}}\mathbf{H}_{k}[n]\mathbf{F}_{\mathrm{RF},i}\mathbf{F}_{\mathrm{BB},i}[n]\mathbf{s}_{i}[n],
\end{equation}
which implies
\begin{equation}
\mathbf{B}(a s_i[n])=a\mathbf{B}(s_i[n]).
\end{equation}
Therefore, we can conclude that the overall operation of beamforming constitutes a linear transformation. 

According to \eqref{eq:receivedsignals}, $\mathbf{Y}_k^{\prime}[n]$ for given $\mathcal{H}_{k}$ can be expressed as
\begin{equation}
\mathbf{Y}_k^\prime[N_\mathrm{c}]=\mathbf{A}_k^\prime[N_\mathrm{c}]\mathbf{s}_k^\prime[N_\mathrm{c}]]+\mathbf{u}_k^\prime[N_\mathrm{c}]],
\end{equation}
where 
$\mathbf{A}_k^\prime[N_\mathrm{c}]$ is the concatenation of $\mathbf{A}_k[n]$, and $\mathbf{A}_k[n]=\mathbf{W}_{\mathrm{BB},k}^{\mathsf{H}}[n]\mathbf{W}_{\mathrm{RF},k}^{\mathsf{H}}\mathbf{H}_k[n]\mathbf{F}_{\mathrm{RF}}\mathbf{F}_{\mathrm{BB},k}[n]$. Note that$\mathbf{A}^{\prime}_k[N_\mathrm{c}]\mathbf{s}_k^{\prime}[N_\mathrm{c}]$ is a affine transformation of $\mathbf{s}_k^{\prime}[N_\mathrm{c}]$, which preserves the properties of the multivariate Gaussian distribution. The term $\mathbf{u}_k^\prime[N_\mathrm{c}]$ is the concatenation of $\tilde{\mathbf{u}}_{k}[n]$, and $\tilde{\mathbf{u}}_{k}[n]=\sum_{i \neq k}\mathbf{W}_{\mathrm{BB},k}^\mathsf{H}[n]\mathbf{W}_{\mathrm{RF},k}^\mathsf{H}\mathbf{H}_{k}[n]\mathbf{F}_{\mathrm{RF}}\mathbf{F}_{\mathrm{BB},i}[n] \mathbf{s}_{i}[n] + \mathbf{W}_{\mathbf{BB},k}^\mathsf{H}[n]\mathbf{W}_{\mathbf{RF},k}^\mathsf{H}\mathbf{u}_{k}[n]$. Similarly, $\mathbf{u}_k^\prime[N_\mathrm{c}]$ also satisfies the multivariate Gaussian distribution. Therefore, $\mathbf{Y}_k^\prime[N_\mathrm{c}]$ follows a multivariate Gaussian distribution.

\section{Derivation of the Transformation Matrix} \label{Transmatrix}
\begin{align}
&\frac{\partial\tau_{\ell_k}}{\partial \mathbf{p}_{\mathrm{r},k}}=\frac{1}{c}\left[\sin\theta_{\mathrm{r},\ell_k}\sin\phi_{\mathrm{r},\ell_k}, \sin\theta_{\mathrm{r},\ell_k}\cos\phi_{\mathrm{r},\ell_k},\cos\theta_{\mathrm{r},\ell_k}\right]^\mathsf{T},\nonumber\\
&\hspace{1cm}\hspace{1cm}\hspace{1cm}\hspace{1cm}\hspace{1cm}\hspace{0.9cm}\text{LoS \& NLoS}\\
&\frac{\partial\phi_{\mathrm{r},\ell_k}}{\partial \mathbf{p}_{\mathrm{r},k}}=\begin{cases}\frac{\left[-\sin\theta_{\mathrm{r},\ell_k},\cos\phi_{\mathrm{r},\ell_k},0\right]^\mathsf{T}}{\|\mathbf{p}_{\mathrm{r},k}-\mathbf{p}_{\mathrm{s},\ell_k}\|_2},\quad \text{NLoS}\\
\frac{\left[-\sin\theta_{\mathrm{r},\ell_k},\cos\phi_{\mathrm{r},\ell_k},0\right]^\mathsf{T}}{\|\mathbf{p}_{\mathrm{r},k}-\mathbf{p}_\mathrm{t}\|_2},\quad \text{LoS}\end{cases}\\
&\frac{\partial\theta_{\mathrm{r},\ell_k}}{\partial \mathbf{p}_{\mathrm{r},k}}=\begin{cases}\frac{1}{\|\mathbf{p}_{\mathrm{r},k}-\mathbf{p}_{\mathrm{s},\ell_k}\|_2}\left[-\cos\phi_{\mathrm{r},\ell_k}\cos\theta_{\mathrm{r},\ell_k},\right.\\
\left.-\sin\phi_{\mathrm{r},\ell_k}\cos\theta_{\mathrm{r},\ell_k},\sin\theta_{\mathrm{r},\ell_k}\right]^\mathsf{T},\text{NLoS}\\
\frac{1}{\|\mathbf{p}_{\mathrm{r},k}-\mathbf{p}_\mathrm{t}\|_2}\left[-\cos\phi_{\mathrm{r},\ell_k}\cos\theta_{\mathrm{r},\ell_k}, \right.\\
\left.-\sin\phi_{\mathrm{r},\ell_k}\cos\theta_{\mathrm{r},\ell_k},\sin\theta_{\mathrm{r},\ell_k}\right]^\mathsf{T},\; \text{LoS}\end{cases}\\
&\frac{\partial\phi_{\mathrm{t},\ell_k}}{\partial \mathbf{p}_{\mathrm{r},k}}=\begin{cases}0,\quad \text{NLoS}\\
\frac{1}{\|\mathbf{p}_{\mathrm{r},k}-\mathbf{p}_\mathrm{t}\|_2}\big[-\sin\theta_{\mathrm{t},\ell_k},\cos\phi_{\mathrm{t},\ell_k},0\big]^\mathsf{T},\quad \text{LoS}\end{cases}\\
&\frac{\partial\theta_{\mathrm{t},\ell_k}}{\partial \mathbf{p}_{\mathrm{r},k}}=\begin{cases}0,\quad \text{NLoS}\\
\frac{1}{\|\mathbf{p}_{\mathrm{r},k}-\mathbf{p}_\mathrm{t}\|_2}\left[-\cos\phi_{\mathrm{t},\ell_k}\cos\theta_{\mathrm{t},\ell_k},\right.\\
\left.-\sin\phi_{\mathrm{t},\ell_k}\cos\theta_{\mathrm{t},\ell_k},\sin\theta_{\mathrm{t},\ell_k}\right]^\mathsf{T},\;\text{LoS}\end{cases}
\end{align}

\section{The proof of lemma (\ref{lemma:J})} \label{appendixJ}
The FIM associated with the $n$-th subcarrier for positioning is defined as $\mathbf{J}_{n}\big(\mathbf{p}_{\mathrm{r},k}\big|\mathcal{H}_{k}\big)$. The PELB of the $n$-th subcarrier is given by 
\begin{equation}
e_{n}(\mathbf{p})=\mathrm{tr}\Big(\mathbf{J}^{-1}_{n}\big(\mathbf{p}_{\mathrm{r},k}\big|\mathcal{H}_{k}\big)\Big).
\end{equation}
For a multi-subcarrier system, the FIM of the received signal matrix $\mathbf{Y}_k^{\prime}[n]$ is the sum of theFIM across all subcarriers:
\begin{equation}
\mathbf{J}\big(\mathbf{p}_{\mathrm{r},k}\big|\mathcal{H}_{k}\big)
=\frac{1}{N_c}\sum_{n=1}^{N_\mathbf{c}}\mathbf{J}_{n}\big(\mathbf{p}_{\mathrm{r},k}\big|\mathcal{H}_{k}\big).
\end{equation}
The MSCPEB of the considered system is
\begin{equation}
e(\mathbf{p})=\mathrm{tr}\Big(\mathbf{J}^{-1}\big(\mathbf{p}_{\mathrm{r},k}\big|\mathcal{H}_{k}\big)\Big)=\mathrm{tr}\Big(\frac{1}{N_c}\sum_{n=1}^{N_\mathrm{c}}\mathbf{J}_{n}\big(\mathbf{p}_{\mathrm{r},k}\big|\mathcal{H}_{k}\big)\Big)^{-1}.
\end{equation}
For positive definite matrices $\mathbf{X}=\mathbf{J}_{n}\big(\mathbf{p}_{\mathrm{r},k}\big|\mathcal{H}_{k}\big)$, the function $f(\mathbf{X}) = \mathrm{tr}(\mathbf{X}^{-1})$ is convex. Using Jensen's inequality, we can obtain the following relationship:
\begin{equation}
\mathrm{tr}\left( \left( \frac{1}{N_c} \sum_{n=1}^{N_c} \mathbf{J}_{n}\big(\mathbf{p}_{\mathrm{r},k}\big|\mathcal{H}_{k}\big) \right)^{-1} \right) \leq \frac{1}{N_c} \sum_{n=1}^{N_c} \mathrm{tr}\big(\mathbf{J}^{-1}_{n}\big(\mathbf{p}_{\mathrm{r},k}\big|\mathcal{H}_{k}\big)\big).
\end{equation}

\section{Derivation of $\left[\mathbf{J}(\eta_{\ell_k}|\mathcal{H}_{k}) \right]_{{\imath},{\jmath}}$} \label{FIMderivatives}

\begin{align}
&\frac{\partial\boldsymbol{\mu}_k[n]}{\partial\tau_{\ell_k}}\nonumber\\
&\hspace{0.5cm}=-j2\pi(f_\mathrm{c}+f_n)\mathbf{W}_{\mathrm{BB},k}^\mathsf{H}[n]\mathbf{W}_{\mathrm{RF},k}^\mathsf{H}\alpha_{\ell_k} e^{-j2\pi(f_\mathrm{c}+f_n)\tau_{\ell_k}}\nonumber\\
&\hspace{0.8cm}  \times \bm{a}_{\mathrm{r},n}(\vartheta_{\mathrm{r},n},\varphi_{\mathrm{r},n})\bm{a}_{\mathrm{t},n}^\mathsf{H}(\vartheta_{\mathrm{t},n},\varphi_{\mathrm{t},n})\mathbf{F}_{\mathrm{RF}}\mathbf{F}_{\mathrm{BB},k}[n], 
\end{align}
\begin{align}
\frac{\partial\bm{\mu}_k[n]}{\partial\theta_{\mathrm{r},{\ell_k}}}&=\mathbf{W}_{\mathrm{BB},k}^\mathsf{H}[n]\mathbf{W}_{\mathrm{RF},k}^\mathsf{H}\frac{\partial\mathbf{H}_k[n]}{\partial\theta_{\mathrm{r},{\ell_k}}}\mathbf{F}_{\mathrm{RF}}\mathbf{F}_{\mathrm{BB},k}[n] \nonumber\\
&=\mathbf{W}_{\mathrm{BB},k}^\mathsf{H}[n]\mathbf{W}_{\mathrm{RF},k}^\mathsf{H}\sqrt{\frac{N_\mathrm{r}N_\mathrm{t}}{L_k}}\sum_{\ell_k=1}^{L_k}\alpha_{\ell_k}e^{-j2\pi(f_\mathrm{c}+f_n)\tau_{\ell_k}} \nonumber\\
\times &\left(\frac{\partial\bm{a}_{\mathrm{r},n}(\vartheta_{\mathrm{r},\ell_k},\varphi_{\mathrm{r},\ell_k})}{\partial\theta_{\mathrm{r},\ell_k}}\right)\bm{a}_{\mathrm{t},n}^{\mathsf{H}}(\vartheta_{\mathrm{r},\ell_k},\varphi_{\mathrm{r},\ell_k})\mathbf{F}_{\mathrm{RF}}\mathbf{F}_{\mathrm{BB},k}[n] \nonumber\\
&=\mathbf{W}_{\mathrm{BB},k}^\mathsf{H}[n]\mathbf{W}_{\mathrm{RF},k}^\mathsf{H}\sqrt{\frac{N_\mathrm{r}N_\mathrm{t}}{L_k}}\sum_{\ell_k=1}^{L_k}\alpha_{\ell_k}e^{-j2\pi(f_\mathrm{c}+f_n)\tau_{\ell_k}} \nonumber\\
&\times \frac{-j2\pi d\sin(\theta_{\mathrm{r},\ell_k})}{\lambda_c}\bm{a}_{\mathrm{r},n}(\vartheta_{\ell_k},\varphi_{\ell_k})\left(-\frac{d\sin(\theta_{\mathrm{r},\ell_k})}{\lambda_c}\right) \nonumber\\
& \times \bm{a}_{\mathrm{t},n}^{\mathsf{H}}(\vartheta_{\mathrm{r},\ell_k},\varphi_{\mathrm{r},\ell_k})\mathbf{F}_{\mathrm{RF}}\mathbf{F}_{\mathrm{BB},k}[n],
\end{align}

\begin{align}
&\frac{\partial\bm{\mu}_k[n]}{\partial\phi_{\mathrm{r},{\ell_k}}} =\mathbf{W}_{\mathrm{BB},k}^\mathsf{H}[n]\mathbf{W}_{\mathrm{RF},k}^\mathsf{H}\frac{\partial\mathbf{H}_k[n]}{\partial\phi_{\mathrm{r},{\ell_k}}}\mathbf{F}_{\mathrm{RF}}\mathbf{F}_{\mathrm{BB},k}[n] \nonumber\\
=&\mathbf{W}_{\mathrm{BB},k}^\mathsf{H}[n]\mathbf{W}_{\mathrm{RF},k}^\mathsf{H}\sqrt{\frac{N_\mathrm{r}N_\mathrm{t}}{L_k}}\sum_{\ell_k=1}^{L_k}\alpha_{\ell_k}e^{-j2\pi(f_\mathrm{c}+f_n)\tau_{\ell_k}} \nonumber\\
&\times \left(\frac{\partial\bm{a}_{\mathrm{r},n}(\vartheta_{\mathrm{r},\ell_k},\varphi_{\mathrm{r},\ell_k})}{\partial\phi_{\mathrm{r},\ell_k}}\right)\bm{a}_{\mathrm{t},n}^{\mathsf{H}}(\vartheta_{\mathrm{r},\ell_k},\varphi_{\mathrm{r},\ell_k})\mathbf{F}_{\mathrm{RF}}\mathbf{F}_{\mathrm{BB},k}[n] \nonumber\\
=&\mathbf{W}_{\mathrm{BB},k}^\mathsf{H}[n]\mathbf{W}_{\mathrm{RF},k}^\mathsf{H}\sqrt{\frac{N_\mathrm{r}N_\mathrm{t}}{L_k}}\sum_{\ell_k=1}^{L_k}\alpha_{\ell_k}e^{-j2\pi(f_\mathrm{c}+f_n)\tau_{\ell_k}} \nonumber\\
\times &\frac{-j2\pi d\cos(\theta_{\mathrm{r},\ell_k})}{\lambda_c}\bm{a}_{\mathrm{r},n}(\vartheta_{\mathrm{r},\ell_k},\varphi_{\mathrm{r},\ell_k})\left(\frac{d\sin(\theta_{\mathrm{r},\ell_k})\cos(\phi_{\mathrm{r},\ell_k})}{\lambda_c}\right) \nonumber\\
& \times \bm{a}_{\mathrm{t},n}^{\mathsf{H}}(\vartheta_{\mathrm{r},\ell_k},\varphi_{\mathrm{r},\ell_k})\mathbf{F}_{\mathrm{RF}}\mathbf{F}_{\mathrm{BB},k}[n],
\end{align}

The same derivation applies to $\frac{\partial\bm{\mu}_k[n]}{\partial\theta_{\mathrm{t},{\ell_k}}}$  and $\frac{\partial\boldsymbol{\mu}_k[n]}{\partial\phi_{\mathrm{t},{\ell_k}}}$.

\section{The derivatives in (\ref{eq:Newton})} \label{sec:Newton}
\vspace{-1.5mm}
\begin{align}&\frac{\partial\Lambda_k[n]}{\partial\eta_{\ell_k,i}}=-2\Re\left\{\left(\mathbf{Y}[n]-\bm{\mu}[n]\right)^\mathsf{H}\frac{\partial \bm{\mu}[n]}{\partial\eta_{\ell_k,i}}\right\}.\\
&\frac{\partial^2\Lambda_k[n]}{\left(\partial\eta_{\ell_k,i}\right)^2} =2\Re\left\{\left(\frac{\partial \bm{\mu}[n]}{\partial\eta_{\ell_k,i}}\right)^\mathsf{H}\frac{\partial \bm{\mu}[n]}{\partial\eta_{\ell_k,i}}\right. \nonumber\\
&\hspace{2.0cm}-(\mathbf{Y}[n]-\bm{\mu}[n])^\mathsf{H}\frac{\partial^{2}\bm{\mu}[n]}{\left(\partial\eta_{\ell_k,i}\right)^{2}}\Bigg\}. 
\end{align}
The derivatives $\frac{\partial \bm{\mu}[n]}{\partial\eta_{\ell_k,i}}$ can be found in Appendix~\ref{FIMderivatives}.

\end{appendices}

\bibliographystyle{IEEEtran} 
\bibliography{references}

\begin{thebibliography}{10}
\providecommand{\url}[1]{#1}
\csname url@samestyle\endcsname
\providecommand{\newblock}{\relax}
\providecommand{\bibinfo}[2]{#2}
\providecommand{\BIBentrySTDinterwordspacing}{\spaceskip=0pt\relax}
\providecommand{\BIBentryALTinterwordstretchfactor}{4}
\providecommand{\BIBentryALTinterwordspacing}{\spaceskip=\fontdimen2\font plus
\BIBentryALTinterwordstretchfactor\fontdimen3\font minus \fontdimen4\font\relax}
\providecommand{\BIBforeignlanguage}[2]{{%
\expandafter\ifx\csname l@#1\endcsname\relax
\typeout{** WARNING: IEEEtran.bst: No hyphenation pattern has been}%
\typeout{** loaded for the language `#1'. Using the pattern for}%
\typeout{** the default language instead.}%
\else
\language=\csname l@#1\endcsname
\fi
#2}}
\providecommand{\BIBdecl}{\relax}
\BIBdecl

\bibitem{8766143}
Z.~Zhang, Y.~Xiao, Z.~Ma, M.~Xiao, Z.~Ding, X.~Lei, G.~K. Karagiannidis, and P.~Fan, ``{6G} wireless networks: Vision, requirements, architecture, and key technologies,'' \emph{IEEE Veh. Technol. Mag.}, vol.~14, no.~3, pp. 28--41, Jul. 2019.

\bibitem{9599656}
H.~Ren, K.~Wang, and C.~Pan, ``Intelligent reflecting surface-aided {URLLC} in a factory automation scenario,'' \emph{IEEE Trans. Commun.}, vol.~70, no.~1, pp. 707--723, Nov. 2022.

\bibitem{9112752}
F.~Hu, Y.~Deng, W.~Saad, M.~Bennis, and A.~H. Aghvami, ``Cellular-connected wireless virtual reality: Requirements, challenges, and solutions,'' \emph{IEEE Commun. Mag.}, vol.~58, no.~5, pp. 105--111, Jun. 2020.

\bibitem{7984759}
M.~Koivisto, A.~Hakkarainen, M.~Costa, P.~Kela, K.~Leppanen, and M.~Valkama, ``High-efficiency device positioning and location-aware communications in dense {5G} networks,'' \emph{IEEE Commun. Mag.}, vol.~55, no.~8, pp. 188--195, Jul. 2017.

\bibitem{8226757}
J.~A. del Peral-Rosado, R.~Raulefs, J.~A. López-Salcedo, and G.~Seco-Granados, ``Survey of cellular mobile radio localization methods: From {1G} to {5G},'' \emph{IEEE Commun. Surv. Tutorials}, vol.~20, no.~2, pp. 1124--1148, Dec. 2018.

\bibitem{10193776}
J.~Zhang, R.~Xi, Y.~He, Y.~Sun, X.~Guo, W.~Wang, X.~Na, Y.~Liu, Z.~Shi, and T.~Gu, ``A survey of mm{W}ave-based human sensing: Technology, platforms and applications,'' \emph{IEEE Commun. Surv. Tutorials}, vol.~25, no.~4, pp. 2052--2087, Jul. 2023.

\bibitem{10054381}
C.-X. Wang, X.~You, X.~Gao, X.~Zhu, Z.~Li, C.~Zhang, H.~Wang, Y.~Huang, Y.~Chen, H.~Haas, J.~S. Thompson, E.~G. Larsson, M.~D. Renzo, W.~Tong, P.~Zhu, X.~Shen, H.~V. Poor, and L.~Hanzo, ``On the road to {6G}: Visions, requirements, key technologies, and testbeds,'' \emph{IEEE Commun. Surv. Tutorials}, vol.~25, no.~2, pp. 905--974, Feb. 2023.

\bibitem{8207426}
I.~A. Hemadeh, K.~Satyanarayana, M.~El-Hajjar, and L.~Hanzo, ``Millimeter-wave communications: Physical channel models, design considerations, antenna constructions, and link-budget,'' \emph{IEEE Commun. Surv. Tutorials}, vol.~20, no.~2, pp. 870--913, Dec. 2018.

\bibitem{6834753}
M.~R. Akdeniz, Y.~Liu, M.~K. Samimi, S.~Sun, S.~Rangan, T.~S. Rappaport, and E.~Erkip, ``Millimeter wave channel modeling and cellular capacity evaluation,'' \emph{IEEE J. Sel. Areas Commun.}, vol.~32, no.~6, pp. 1164--1179, Jun. 2014.

\bibitem{10422712}
Q.~Xue, C.~Ji, S.~Ma, J.~Guo, Y.~Xu, Q.~Chen, and W.~Zhang, ``A survey of beam management for mmwave and {TH}z communications towards {6G},'' \emph{IEEE Commun. Surv. Tutorials}, vol.~26, no.~3, pp. 1520--1559, Feb. 2024.

\bibitem{3gpp2019nr}
``Study on nr positioning support tr 38.855,'' 3GPP, Sophia Antipolis, France, Technical Report TR 38.855, 2019.

\bibitem{9665433}
S.~Bartoletti, H.~Wymeersch, T.~Mach, O.~Brunnegård, D.~Giustiniano, P.~Hammarberg, M.~F. Keskin, J.~O. Lacruz, S.~M. Razavi, J.~Rönnblom, F.~Tufvesson, J.~Widmer, and N.~B. Melazzi, ``Positioning and sensing for vehicular safety applications in {5G} and beyond,'' \emph{IEEE Commun. Mag.}, vol.~59, no.~11, pp. 15--21, Dec. 2021.

\bibitem{9815098}
R.~Zhang, L.~Cheng, S.~Wang, Y.~Lou, W.~Wu, and D.~W.~K. Ng, ``Tensor decomposition-based channel estimation for hybrid mm{W}ave massive {MIMO} in high-mobility scenarios,'' \emph{IEEE Trans. Commun.}, vol.~70, no.~9, pp. 6325--6340, Jul. 2022.

\bibitem{6799306}
S.~Jeong, O.~Simeone, A.~Haimovich, and J.~Kang, ``Beamforming design for joint localization and data transmission in distributed antenna system,'' \emph{IEEE Trans. Veh. Technol.}, vol.~64, no.~1, pp. 62--76, Apr. 2015.

\bibitem{10323191}
H.~Xv, Y.~Sun, Y.~Zhao, M.~Peng, and S.~Zhang, ``Joint beam scheduling and beamforming design for cooperative positioning in multi-beam leo satellite networks,'' \emph{IEEE Trans. on Veh. Technol.}, vol.~73, no.~4, pp. 5276--5287, Nov. 2024.

\bibitem{10547696}
X.~Chu, Z.~Lu, J.~Kang, Y.~Zou, H.~Zhang, and X.~Qiu, ``Hybrid beamforming toward positioning enhancement under cellular {MIMO} systems,'' \emph{IEEE Trans. Wireless Commun.}, vol.~23, no.~10, pp. 13\,545--13\,561, Jun. 2024.

\bibitem{10579914}
H.~Li, Z.~Wang, X.~Mu, P.~Zhiwen, and Y.~Liu, ``Near-field integrated sensing, positioning, and communication: A downlink and uplink framework,'' \emph{IEEE J. Sel. Areas Commun.}, vol.~42, no.~9, pp. 2196--2212, Jul. 2024.

\bibitem{9578931}
M.~Koivisto, J.~Talvitie, E.~Rastorgueva-Foi, Y.~Lu, and M.~Valkama, ``Channel parameter estimation and {TX} positioning with multi-beam fusion in {5G} mm{W}ave networks,'' \emph{IEEE Trans. Wireless Commun.}, vol.~21, no.~5, pp. 3192--3207, Oct. 2022.

\bibitem{9566601}
S.~Fan, W.~Ni, H.~Tian, Z.~Huang, and R.~Zeng, ``Carrier phase-based synchronization and high-accuracy positioning in {5G} new radio cellular networks,'' \emph{IEEE Trans. Commun.}, vol.~70, no.~1, pp. 564--577, Oct. 2022.

\bibitem{8709732}
A.~R. Finelli and Y.~Bar-Shalom, ``Detection, location estimation, and {CRLB} of a streaking target in an {FPA} with a poisson model,'' \emph{IEEE Trans. Aerosp. Electron. Syst.}, vol.~56, no.~1, pp. 356--367, May 2020.

\bibitem{hua}
H.~Hua, J.~Xu, and R.~Zhang, ``Near-field integrated sensing and communication with extremely large-scale antenna array,'' \emph{arXiv preprint arXiv: 2407.17237}, 2024.

\bibitem{10050406}
Z.~Wang, X.~Mu, and Y.~Liu, ``{STARS} enabled integrated sensing and communications,'' \emph{IEEE Trans. Wireless Commun.}, vol.~22, no.~10, pp. 6750--6765, Feb. 2023.

\bibitem{su2025joint}
T.~Su, Z.~Zhang, J.~Chu, R.~Zhao, and Y.~Huang, ``Joint {DOA} and {TOA} estimation for moving {MIMO} radar based on {OFDM-ISAC} signals,'' \emph{IEEE Trans. on Veh. Technol.}, pp. 1--16, Feb 2025.

\bibitem{tse2005fundamentals}
D.~Tse and P.~Viswanath, \emph{Fundamentals of Wireless Communication}.\hskip 1em plus 0.5em minus 0.4em\relax Cambridge, U.K.: Cambridge Univ. Press, 2005.

\bibitem{wang2020optimal}
T.~Wang, S.~C. Liew, and S.~S. Ullah, ``Optimal rate-diverse wireless network coding over parallel subchannels,'' \emph{IEEE Trans. on Commun.}, vol.~68, no.~8, pp. 4891--4904, Aug 2020.

\bibitem{10370743}
J.~Lin, Y.~Jing, and X.~Yu, ``Cram\'{e}r-rao lower bound analysis of positioning with planar large intelligent surfaces under rician channel,'' \emph{IEEE Trans. Wireless Commun.}, vol.~23, no.~7, pp. 7668--7682, Dec. 2024.

\bibitem{9011751}
B.~Zhou, A.~Liu, V.~Lau, J.~Wen, S.~Mumtaz, A.~K. Bashir, and S.~H. Ahmed, ``Performance limits of visible light-based positioning for internet-of-vehicles: Time-domain localization cooperation gain,'' \emph{IEEE Trans. Intell. Transp. Syst.}, vol.~22, no.~8, pp. 5374--5388, Feb. 2021.

\bibitem{10702556}
S.~Singh, S.~Kumar, S.~Majhi, U.~Satija, and C.~Yuen, ``Blind carrier frequency offset estimation techniques for next-generation multicarrier communication systems: {c}hallenges, comparative analysis, and future prospects,'' \emph{IEEE Commun. Surv. Tutorials}, vol.~27, no.~1, pp. 1--36, Oct. 2025.

\bibitem{boyd2004convex}
S.~Boyd and L.~Vandenberghe, \emph{Convex Optimization}.\hskip 1em plus 0.5em minus 0.4em\relax Cambridge University Press, 2004.

\bibitem{8804387}
A.~Kakkavas, M.~H. Castañeda~García, R.~A. Stirling-Gallacher, and J.~A. Nossek, ``Performance limits of single-anchor millimeter-wave positioning,'' \emph{IEEE Trans. Wireless Commun.}, vol.~18, no.~11, pp. 5196--5210, Aug. 2019.

\bibitem{7397861}
X.~Yu, J.-C. Shen, J.~Zhang, and K.~B. Letaief, ``Alternating minimization algorithms for hybrid precoding in millimeter wave {MIMO} systems,'' \emph{IEEE J. Sel. Top. Signal Process.}, vol.~10, no.~3, pp. 485--500, Feb. 2016.

\bibitem{wang2024performance}
Z.~Wang, X.~Mu, and Y.~Liu, ``Performance analysis of near-field sensing in wideband {MIMO} systems,'' \emph{arXiv preprint arXiv:2404.05076}, 2024.

\bibitem{affes1998new}
S.~Affes and P.~Mermelstein, ``A new receiver structure for asynchronous cdma: Star-the spatio-temporal array-receiver,'' \emph{IEEE J. Sel. Areas Commun.}, vol.~16, no.~8, pp. 1411--1422, Oct 1998.

\bibitem{feder1988parameter}
M.~Feder and E.~Weinstein, ``Parameter estimation of superimposed signals using the {EM} algorithm,'' \emph{IEEE Trans. Acoust., Speech Signal Process.}, vol.~36, no.~4, pp. 477--489, Apr 1988.

\bibitem{2004MitsubishiER}
S.~Gezici and Z.~Sahinoglu, ``{UWB} geolocation techniques for {IEEE} 802.15.4a personal area networks,'' MERL Technical Report, Cambridge, MA, Tech. Rep., Aug. 2004.

\bibitem{10063228}
P.~Wu, J.~Chen, S.~Guo, G.~Cui, L.~Kong, and X.~Yang, ``{NLOS} positioning for building layout and target based on association and hypothesis method,'' \emph{IEEE Trans. Geosci. Remote Sens.}, vol.~61, pp. 1--13, Mar. 2023.

\bibitem{9834216}
K.~Liu, Z.~Tian, Z.~Li, and X.~Wan, ``A non-line-of-sight wireless indoor localization system using custom-designed radio relay and uniform circular array,'' \emph{IEEE Trans. Antennas Propag.}, vol.~70, no.~11, pp. 11\,045--11\,058, Jul. 2022.

\bibitem{9152004}
Z.~Li, Z.~Tian, Z.~Wang, and Z.~Zhang, ``Multipath-assisted indoor localization using a single receiver,'' \emph{IEEE Sens. J.}, vol.~21, no.~1, pp. 692--705, Jul. 2021.

\bibitem{8515231}
R.~Mendrzik, H.~Wymeersch, G.~Bauch, and Z.~Abu-Shaban, ``Harnessing {NLOS} components for position and orientation estimation in {5G} millimeter wave {MIMO},'' \emph{IEEE Trans. Wireless Commun.}, vol.~18, no.~1, pp. 93--107, Oct. 2019.

\bibitem{10741287}
L.~Guo, T.~Lv, and J.~Zeng, ``Angle-based positioning estimation leveraging diffuse scattering paths in millimeter-wave {MIMO} systems,'' \emph{IEEE Sensors Journal}, vol.~24, no.~24, pp. 41\,597--41\,609, Nov. 2024.

\bibitem{9032328}
H.~Kim, K.~Granström, L.~Gao, G.~Battistelli, S.~Kim, and H.~Wymeersch, ``{5G} {mmWave} cooperative positioning and mapping using multi-model {PHD} filter and map fusion,'' \emph{IEEE Trans. Wireless Commun.}, vol.~19, no.~6, pp. 3782--3795, Mar. 2020.

\bibitem{TR38901}
3GPP, ``Study on channel model for frequencies from 0.5 to 100 {GH}z (release 15),'' 3GPP, Tech. Rep. TR 38.901, 2017.

\bibitem{Jakes1993}
W.~Jakes, \emph{Microwave Mobile Communications}.\hskip 1em plus 0.5em minus 0.4em\relax Microwave Mobile Communications, 1993.

\bibitem{10988576}
Z.~Wang, X.~Mu, and Y.~Liu, ``Performance analysis of near-field sensing in wideband {MIMO} systems,'' \emph{IEEE Trans. on Wireless Commun.}, Early Access, Mar. 2025.

\bibitem{3gpp2019tr38855}
{3GPP}, ``Study on {NR} positioning support (release 16),'' 3GPP, Tech. Rep. TR 38.855, 2019.

\end{thebibliography}

\end{document}